\documentclass[journal]{IEEEtran}

\usepackage{moreverb}

\usepackage[colorlinks,citecolor=red,urlcolor=red]{hyperref}

\usepackage{algorithm, algorithmic}
\usepackage{bm}
\usepackage{cite}
\usepackage{graphicx}
\usepackage{amsmath}
\usepackage{amstext,amsxtra,amssymb,underscore}
\usepackage{accents}

\usepackage{subfigure}
\usepackage{caption2}

\usepackage{color}

\makeatletter

\newcommand{\Rmnum}[1]{\expandafter\@slowromancap\romannumeral #1@}
\makeatother

\makeatletter
\def\wideubar{\underaccent{{\cc@style\underline{\mskip8mu}}}}
\def\Wideubar{\underaccent{{\cc@style\underline{\mskip6mu}}}}
\makeatother

\makeatletter
\def\widebar{\accentset{{\cc@style\underline{\mskip8mu}}}}
\def\Widebar{\accentset{{\cc@style\underline{\mskip6mu}}}}
\makeatother

\newtheorem{theorem}{Theorem}
\newtheorem{assumption}{Assumption}
\newtheorem{lemma}{Lemma}
\newtheorem{definition}{Definition}
\newtheorem{remark}{Remark}
\newtheorem{corollary}{Corollary}
\newtheorem{proposition}{Proposition}

\newcounter{MYtempeqncnt}

\usepackage{amstext,amsxtra,amssymb,underscore}
\usepackage{amssymb,bm}

\def\bmg{{\ensuremath{\bm{g}}}}
\def\bmh{{\ensuremath{\bm{h}}}}

\def\bmx{{\ensuremath{\bm{x}}}}
\def\bmy{{\ensuremath{\bm{y}}}}

\def\bmA{{\ensuremath{\bm{A}}}}

\def\bmC{{\ensuremath{\bm{C}}}}
\def\bmD{{\ensuremath{\bm{D}}}}
\def\bmE{{\ensuremath{\bm{E}}}}
\def\bmF{{\ensuremath{\bm{F}}}}
\def\bmG{{\ensuremath{\bm{G}}}}

\def\bmI{{\ensuremath{\bm{I}}}}
\def\bmJ{{\ensuremath{\bm{J}}}}
\def\bmK{{\ensuremath{\bm{K}}}}
\def\bmL{{\ensuremath{\bm{L}}}}
\def\bmM{{\ensuremath{\bm{M}}}}
\def\bmN{{\ensuremath{\bm{N}}}}

\def\bmP{{\ensuremath{\bm{P}}}}
\def\bmQ{{\ensuremath{\bm{Q}}}}
\def\bmR{{\ensuremath{\bm{R}}}}
\def\bmS{{\ensuremath{\bm{S}}}}
\def\bmT{{\ensuremath{\bm{T}}}}
\def\bmU{{\ensuremath{\bm{U}}}}
\def\bmV{{\ensuremath{\bm{V}}}}
\def\bmW{{\ensuremath{\bm{W}}}}
\def\bmX{{\ensuremath{\bm{X}}}}
\def\bmY{{\ensuremath{\bm{Y}}}}
\def\bmZ{{\ensuremath{\bm{Z}}}}
\def\bmomega{{\ensuremath{\bm{\omega}}}}
\def\bmupsilon{{\ensuremath{\bm{\upsilon}}}}

\def\ccal0{{\ensuremath{\mathcal 0}}}
\def\bbx{{\ensuremath{\mathbf x}}}
\def\bby{{\ensuremath{\mathbf y}}}

\def\bb0{{\ensuremath{\mathbf 0}}}

\makeatletter
\def\wideubar{\underaccent{{\cc@style\underline{\mskip8mu}}}}
\def\Wideubar{\underaccent{{\cc@style\underline{\mskip6mu}}}}
\makeatother

\makeatletter
\def\widebar{\accentset{{\cc@style\underline{\mskip8mu}}}}
\def\Widebar{\accentset{{\cc@style\underline{\mskip6mu}}}}
\makeatother

\begin{document}

\newenvironment{centergpic}{}{\begin{center}~\box\graph~\end{center}}



\title{{\color{black}{Power Scheduling of Kalman Filtering in Wireless Sensor Networks with Data Packet Drops}}
}

\author{Gang~Wang,
~\IEEEmembership{Student~Member,~IEEE,}~Jie~Chen,~\IEEEmembership{Senior~Member,~IEEE,}\\Jian~Sun,~\IEEEmembership{Member,~IEEE},~and~Yongjian Cai
\thanks{This work was supported in part by the Natural Science Foundation of China under Grant 61104097, National Science Foundation for
Distinguished Young Scholars of China under Grant 60925011, Projects of Major International (Regional) Joint Research Program NSFC under Grant 61120106010, Beijing Education Committee Cooperation Building Foundation Project XK100070532, and Research Fund for the Doctoral Program of Higher Education of China 20111101120027. G. Wang was supported in part by
the China Scholarship Council.}
\thanks{The authors are with the School of Automation, Beijing Institute of Technology,
Beijing 100081, China, and also with Key Laboratory of Intelligent Control and Decision of Complex Systems. 
(E-mail: wang4937@umn.edu, chenjie@bit.edu.cn, sunjian@bit.edu.cn, caiyongjian@bit.edu.cn).

The corresponding author of this paper is J. Sun.

}
}

\maketitle

\begin{abstract}
For a wireless sensor network (WSN) with a large number of low-cost, battery-driven, multiple transmission power leveled sensor nodes of limited transmission bandwidth, then conservation of transmission resources (power and bandwidth) is of paramount importance. Towards this end, this paper considers the problem of power scheduling of Kalman filtering for general linear stochastic systems subject to data packet drops (over a packet-dropping wireless network). The transmission of the acquired measurement from the sensor to the remote estimator is realized by sequentially transmitting every single component of the measurement to the remote estimator in one time period. The sensor node decides separately whether to use a high or low transmission power to communicate every component to the estimator across a packet-dropping wireless network based on the rule that promotes the power scheduling with the least impact on the estimator mean squared error. 
{\color{black}{Under the customary assumption that the predicted density is (approximately) Gaussian, leveraging the statistical distribution of sensor data, the mechanism of power scheduling, the wireless network effect and the received data, the minimum mean squared error estimator is derived. By investigating the statistical convergence properties of the estimation error covariance, we establish, for general linear systems, both the sufficient condition and the necessary condition guaranteeing the stability of the estimator.}}
\end{abstract}


%
%

\section{Introduction}
\label{section 1}
With the groundbreaking advances of microsensor technology and wireless communication technology, wireless sensor
networks (WSNs) have been found in a plethora of applications. The proposed and/or already deployed applications include, but not limited to, battlefield surveillance, intelligent transportation systems, health care, environment monitoring and control, disaster prevention and recovery, and more efficient electric power grids \cite{cn2002su, tsp2006ribeiro1, tsp2006ribeiro2, tsp2006ribeiro, tsp2008msechu, tsp2012msechu, tac2012jia}. However, there are still some severe limitations in current WSNs that prevent them from better serving the people, such as, limited power at each battery-driven sensor, limited communication ability, limited computation ability and limited wireless bandwidth \cite{tac2012jia}. These limitations will ineluctably bring some challenging problems to the study of estimation and control over WSNs. Therefore, it is of great significance to investigate how to conserve transmission power and bandwidth while achieving a similar estimation performance.

Towards this end, recurring attention has been paid to the research of remote estimation under communication resources (energy constraint and bandwidth) requirement in the last decade and a multitude of publications can be widely found in the literature; see, for example, \cite{tsp2012you, tsp2006ribeiro1,tsp2006ribeiro2,cs2010ribeiro,tsp2007schizas,tsp2008msechu,tsp2012msechu,tit2005luo,tsp2006luo,tac2012jia,tac2009savage,
spm2006luo,tsp2006ribeiro,auto2007suh,timc2011you,tsp2012shixie,auto2012battistelli,tac2013shi,auto2011shi,tsp2011yang,tsp2013you,iet2013wang, ccc2012you} 
and references therein. 
Among them, by the desire of conserving transmission energy and bandwidth, various methods regarding measurement quantization, censoring, and dimensionality-reduction were specialized in \cite{tsp2006ribeiro, tsp2006ribeiro1, tsp2006ribeiro2, cs2010ribeiro, tsp2007schizas, tsp2008msechu, tsp2012msechu,
tsp2006luo, spm2006luo, timc2011you}. 
Another creative method in terms of measurement scheduling has been extensively studied in \cite{tsp2012you, tac2009savage, auto2007suh, tsp2012shixie, auto2012battistelli, tac2013shi, auto2011shi, tsp2011yang, tsp2013you} etc.

 To be more specific, owing to the power-limited nature of wireless sensors and the fact that replacing the exhausted batteries are costly operations and may even be impossible, only a limited number of measurement transmissions can thereby be made by the wireless devices in most WSNs applications. In \cite{tac2009savage}, optimal measurement scheduling policies were devised for a particular class of scalar Gauss-Markov systems to minimize the terminal estimation error variance over a given time horizon $T$, in which only $p<T$ measurements can be taken and transmitted to the remote estimator side. In practical, most commercially available sensor nodes nowadays have multiple transmission power levels \cite{tsp2006luo} and it is assumed that high transmission power leads to reliable data flow while low transmission power may cause unreliable data flow \cite{auto2011shi} and therefore data packet drops may occur. {\color{black}The results in \cite{tac2009savage} were then recently extended to a special class of high-order Gauss-Markov systems in \cite{tsp2012shixie}, where both the sensor energy constraint and data packet drops were taken into account and furthermore, two scenarios in terms of sensor nodes with limited or sufficient computation capacity are considered. Under some appropriate conditions, the optimal schedulers derived indicate that the $p$ measurement transmissions should be distributed along the last $p$ time steps over the time horizon $T,$ that is, from $T-1-p$ to $T-1.$ It is worth noticing that the optimal measurement schedulers above are deterministic, which are so-called ``offline schedulers,''  and therefore, this kind of offline schedulers have the apparent advantage of offline determination of optimal scheduling schemes. Nevertheless, also noticed that the estimation error covariance matrix increases drastically for unstable systems in the first $T-p$ time steps owing to no measurements transmitted to the estimator side to update the covariance prediction, which is a disadvantage of these offline schedulers. More discussions and generalizations on offline schedulers can also be found in \cite{auto2011shi} and \cite{tsp2011yang}.}

 On the other hand, to avoid the disadvantage mentioned above, schedulers taking the current measurement value into consideration were devised in \cite{tsp2012you,auto2007suh,auto2012battistelli,tsp2013you,tac2013shi} and considering the modified Kalman filter therein is very much involved with a stochastic variable, these schedulers are called ``online schedulers.'' The send-on-delta strategy was adopted in \cite{auto2007suh} to reduce sensor data traffic by transmitting sensor data only if their values change exceeds a prescribed threshold. However, the threshold has no analytic relationship with the estimation performance and no stability and performance analysis were given with respect to the proposed modified Kalman filter. Innovation-based measurement schedulers were primarily constructed in \cite{tsp2012you}, \cite{tac2013shi} by quantifying the ``importance'' of every measurement using the normalized measurement innovations. The main idea is that only ``important" enough measurements will be transmitted to the estimator side to update the state prediction and covariance prediction, and when the transmission does not occur, the additionally known information based on given threshold of the scheduler will be utilized. Moreover, some stability analysis of Kalman filtering with the aforementioned two stochastic schedulers was presented in \cite{tsp2012you} and {\color{black}{however, only necessary conditions guaranteeing the convergence of expected estimation error covariance were established for systems with full-row-ranked observation matrix therein.}}

Inspired by those observations, this paper builds on and considerably broads the scope of \cite{tsp2012shixie} and \cite{tsp2012you}, where the power scheduler is dependent on the time-horizon $T$ and the covariance increases drastically during the first $T-p-1$ time steps. 
{\color{black}In comparison, the main contributions of this work are twofold and summarized as follows.
\begin{enumerate}
\item We consider power scheduling problem of remote state estimation of general high-order linear stochastic systems. Data packet drop, a typical and natural phenomenon in wireless networks, is also considered and modeled as one Bernoulli i.i.d. process. See Fig. \ref{fig1} for an illustration, where the power scheduler is embedded in the sensor node. We devise a component-wise innovation-based power scheduler and the corresponding minimum mean squared error estimator (MMSE).

\item We investigate the statistical convergence properties of the estimation error covariance matrix by constructing one auxiliary function and we establish both the sufficient condition and the necessary condition for convergence of the averaged estimation error covariance. Theorem \ref{sufficient condition} originally establishes the sufficient condition for mean square stability of estimation error covariance matrix and Theorem \ref{necessary condition} extends the results for systems with full-row-ranked observation matrix in the literature to general linear systems.  Therefore, this work is an important generalization of and a necessary complementary to the literature of state estimation of WSNs in the sense of both estimation framework and theoretical stability analysis; see, for example, \cite{tsp2012you, tsp2006ribeiro, cs2010ribeiro, tac2004sinopoli, cdc2007garone, auto2011youxie}.
%

\end{enumerate}

\begin{figure}
\begin{minipage}[b]{0.5\textwidth}
\centering
\includegraphics[scale=0.6]{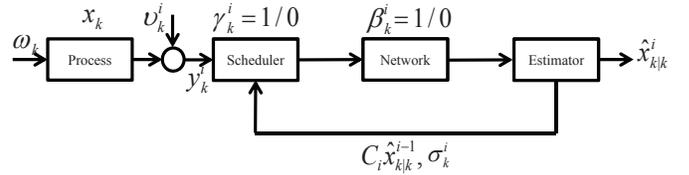}
{\color{black}{\caption{Network architecture.}}}
\label{fig1}
\end{minipage}%
\end{figure}
}

Coincidentally, from a mathematical point of view, the stability analysis can be cast into the well-received category of Kalman filtering with incomplete (dropped or delayed) observations primarily studied in \cite{tac2004sinopoli, cdc2004goldsmith, cdc2007garone, tac2008luca, auto2009shixie} and lately in \cite{tac2012sinopoli, arxiv2009sinopoli}. Explicit comparisons made between the present work and those pioneering works definitely show the implications and necessity of this work. Part of the material in this paper was presented in \cite{cyber2013wang}.

 The remainder of this paper is organized as follows. We briefly introduce the measurement model as well as some standard assumptions in Section \ref{section 2} and devise the minimum mean squared error estimator with power scheduler in Section \ref{section 3}. In Section \ref{section c}, we provide both sufficient condition and necessary condition that guarantee the convergence of averaged estimation error covariance matrix. Finally, conclusions and current research threads are outlined in Section \ref{section 4}.

\emph{Notations:}
{\color{black}{Straight boldface denote the multivariate quantities such as vectors (lowercase) and matrices (uppercase). Let $Q(\cdot)$ be the tail probability of the standard normal distribution, i.e., $Q(x)=1/\sqrt{2\pi}\cdot\int_x^{\infty} {\rm{exp}}(-t^2/2)dt.$ Denote $\bmx\sim\mathcal{N}(\bm{\mu},\bm{\Sigma})$ or $\mathcal{N}(\bmx;\bm{\mu},\bm{\Sigma})$ by a normally distributed vector $\bmx$ with mean $\bm{\mu}$ and covariance $\bm{\Sigma}.$ For random vectors $\bmx$ and $\bmy,$ $E[\bmx]$ denotes the expectation value of $\bmx,$ and $\bmx|\bmy$ denotes the conditional random vector when $\bmy$ is given. Furthermore, we use $(\bm{\cdot})'$ to denote the transpose of a matrix, use $\bmP>{\bm0}~(\ge{\bm0})$ to represent the positive definite (positive semi-definite) matrix $\bmP,$ and use ${\rm{diag}}\{l_1,l_2,\ldots,l_m\}$ to denote the diagonal matrix with the main diagonal elements $l_1,l_2,\ldots,l_m.$ $\bmI_n$ denotes the $n\times n$ identity matrix and ${\bm0}$ denotes the zero matrix of appropriate dimensions. The notation $\otimes$ denotes the Kronecker product of two matrices. The mean square stability of the filter, i.e., ${\rm{sup}}_{k\in N}\mathbb{E}[\bmP_k]< \infty,$ implies there always exists a positive definite matrix $\widebar{\bmP}$ such that $\bmP_k\leq \widebar{\bmP}$ for all $k\in N$ \cite{auto2011youxie}, where the mathematical expectation is taken with respect to both the random power scheduling process and random packet drop process in this paper. For two positive definite matrices $\bmP$ and $\bmQ,$ the matrix inequality $\bmP\geq \bmQ$ means matrix $\bmP-\bmQ$ is positive semidefinite. Similar notations will be made for $\bmP\leq \bmQ, \bmP> \bmQ$ and $\bmP< \bmQ.$}}


\section{Problem Formulation and Preliminaries}
\label{section 2}
Consider the following linear discrete-time stochastic system:
\begin{align}%
\bmx_{k+1}&=\bmA\bmx_{k}+\bmomega_{k}\label{system}\\
\bmy_{k}&=\bmC\bmx_{k}+\bmupsilon_{k}\label{measurement}
\end{align}
where $\bmx_{k}\in\mathbb{R}^{n}$ is the state vector and $\bmy_{k}\in\mathbb{R}^{m}$ is the measurement vector, $\bmomega_{k}\in\mathbb{R}^{n}$ and $\bmupsilon_{k}\in\mathbb{R}^{m}$ are Gaussian random vectors with zero-means and covariance matrices $\bmQ\ge {\bm0}$ and $\bmR>{\bm0},$ respectively. The initial state $\bmx_0$ is also assumed to be a Gaussian random vector with mean $\hat{\bm{x}}_0$ and covariance matrix $\bmP_0>{\bm0}.$ It is further posited that the random vectors $\bmomega_k,\bmupsilon_k,\bmx_0$ are mutually independent.

{\color{black}{ We assume a high transmission energy leads to reliable data flow while a low transmission energy may result in data packet drops during wireless network communications. This assumption is reasonable and motivated by the two facts: Most economically available sensors in the market have multiple transmission energy levels to choose from \cite{tsp2006luo} and higher transmission energy leads to a higher signal-to-noise ratio (SNR) at the remote estimator, which can be simply interpreted as a higher packet arrival rate \cite{book2007}.}} Therefore, once communication failure occurs, the whole data packet will drop. {\color{black}For simplicity, the present paper considers that the sensor node has only two transmission power levels \cite{auto2011shi}, \cite{tsp2012shixie} and, though, results derived in this paper can be easily generalized to multiple transmission power level case. Specifically, when a high transmission power $\Delta$ is employed, the data packet can be successfully delivered to the estimator side; when a low transmission power $\delta$ is employed, then the data packet is supposed to arrive at the estimator side only with a probability $\beta\in(0,1)$. Similar power scheduling has also been considered in \cite{tsp2012shixie} with a different estimation framework.}


Before delving into the mechanism of power scheduling, the following two standard assumptions are presented.
\begin{assumption}%
\label{assumption 4}
$\left(\bmA,\bmQ^{1/2}\right)$ is controllable and $(\bmC,\bmA)$ is observable.
\end{assumption}
\begin{assumption}
\label{assumption 1}
The covariance $\bmR$ is diagonal, i.e., $\bmR={\rm{diag}}\{R_{1},R_{2},\ldots,R_{m}\}$.
\end{assumption}

{\color{black}\begin{remark}
In fact, if the measurement noise vectors $\bmupsilon_k$ are white, then covariance matrix $\bmR$ is diagonal. If $\bmupsilon_k$ are not white and $\bmR$ is thus a general positive definite matrix, the idea is primarily to whiten the observations. To this end, we define the square root matrix of a positive definite matrix $\bmR$ as $\bmR:=\bmR^{1/2}\left(\bmR^{1/2}\right)'.$ Instead of using $\bmy_k=\bmC\bmx_k+\bmupsilon_k,$ we consider a transformed measurement
\begin{align*}
\tilde{\bmy}_k:=&\bmR^{-1/2}\bmy_k\\
=&\bmR^{-1/2}\bmC\bmx_k+\bmR^{-1/2}\upsilon_k\\
:=&\tilde{\bmC}\bmx_k+\tilde{\bmupsilon}_k
\end{align*}
where $\mathbb{E}\left[\tilde{\bmupsilon}_k\tilde{\bmupsilon}_k'\right]=\bmI_n.$ Therefore without loss of generality, we can assume the measurement noise covariance $\bmR$ to be diagonal.
\end{remark}}

\section{Power Scheduling and \\Sequential Kalman Filtering}
\label{section 3}
{\color{black}{In this paper, a round-robin, slotted-time measurement transmission policy is envisioned such that, only a scalar is allowed to be communicated to the estimator at every transmission and one sampling interval (i.e., one time instant from each $k$ to $k+1$) can be explicitly partitioned into $m$ (the dimension of measurement vector $\bmy_k$) time slots, and at the $i$th time slot $T_i,$ the scheduler located at the sensor node decides whether to use the high or low transmission energy to transmit $y_k^i,$ the $i$th component of $\bmy_k.$  This sensor scheduling protocol was also used in \cite{tsp2012msechu}.}} 

It is well acknowledged that the measurement innovation indicates new information of the current measurement that is not contained in all historical measurements and intuitively speaking, a large innovation represents the current measurement is quite different than the predicted measurement and therefore contains much useful information to update the estimate. Thus, we define the measurement of large innovation as ``important'' measurement and otherwise, less ``important'' measurement. In this sense, we devise an innovation-based power scheduling policy, which compares the normalized measurement innovation with a given threshold to quantify the ``importance'' of every measurement and then uses a high (or low) transmission power to communicate the ``important'' (or less ``important'') measurement.

Specifically, at time instant $k,$ let the binary random variables $\gamma_k^i~(0\text{~or~}1),$ $i=1,2,\ldots,m,$ represent whether the transmission power $\Delta$ or $\delta$ is utilized for transmission of $y_k^i.$ Let another sequence of random variables $\beta_k^i=1$ or $0$ for $i=1,2,\ldots,m,$ indicate whether the data packet $y_k^i$ arrives at estimator side successfully or not. {\color{black}{Throughout this paper, we postulate that the values of $\gamma_k^i$ and $\beta_k^i$ for all $i=1,2,\ldots,m$ at every $k$ can be observed; since we can employ TCP-like protocols where the packet acknowledgements are guaranteed at every time instant to notify estimator whether the data packet is received \cite{tac2012sinopoli, cdc2007garone}.}} For future reference, define $\mathcal{I}_k^i=\big\{\gamma_1^1y_1^1,(1-\gamma_1^1)\beta_1^1y_1^1,\gamma_1^1,(1-\gamma_1^1)\beta_1^1,
\gamma_1^2y_1^2,(1-\gamma_1^2)\beta_1^2y_1^2,\gamma_1^2,(1-\gamma_1^2)\beta_1^2,\ldots,\gamma_k^iy_k^i,
(1-\gamma_k^i)\beta_k^iy_k^i,\gamma_k^i,(1-\gamma_k^i)\beta_k^i\big\},$ and $\mathcal{G}_k^i=\left\{\mathcal{I}_k^{i-1},y_k^i\right\},$
$i=1,2,\ldots,m.$

{\color{black}{For any time instant $k,$ denote by $\hat{\bmx}_{k|k}^i$ the mean squared error estimate of $x_k$ at estimator based upon all received information at the end of $i$th time slot and likewise, by $\bmP_{k|k}^i$ the estimation error covariance, i.e.,
\begin{align}
\hat{\bmx}_{k|k}^i&=\mathbb{E}\left[\bmx_k\big|\mathcal{I}_k^{i}\right],\\
\bmP_{k|k}^i&=\mathbb{E}\left[(\bmx_k-\hat{\bmx}_{k|k}^{i})(\bmx_k-\hat{\bmx}_{k|k}^{i})'\big|
\mathcal{I}_{k}^{i}\right].
\end{align}
Let $\bmy_k'=[y_k^1,y_k^2,\ldots,y_k^m]$ {\color{black} {and $\eta_i,~i=1,2,\ldots,m,$ be given fixed thresholds.}} Then the developed power scheduler and the corresponding MMSE estimator are together tabulated as Algorithm \ref{ps}.}}

\begin{algorithm}
\caption{(Local Power Scheduler and Remote MMSE Estimator)}\label{ps}
\begin{algorithmic}
\STATE {\textbf{Initialization:}\\
$\hat{\bmx}_{0|0}=\hat{\bmx}_{0}, \bmP_{0|0}=\bmP_{0}$,}
\STATE {\textbf{Time prediction:} given $\hat{\bmx}_{k-1|k-1}$ and $\bmP_{k-1|k-1}$, do\\
$\hat{\bmx}_{k|k-1}=\bmA\hat{\bmx}_{k-1|k-1}, \bmP_{k|k-1}=\bmA\bmP_{k-1|k-1}\bmA'+\bmQ.$\\
\STATE {\textbf{Power scheduling, measurement transmission and measurement update:}}\\
Define $\bmy_{k}'=\left[y_{k}^{1},y_k^2,\ldots,y_{k}^{m}\right],$ $\bmC'=\left[\bmC_{1}',\bmC_2',\ldots,\bmC_{m}'\right],$\\ $\hat{\bmx}_{k|k}^{0}=\hat{\bmx}_{k|k-1}$ and $\bmP_{k|k}^{0}=\bmP_{k|k-1}.$\\
For $i=1,2,\ldots,m,$ set\\
{\color{black}{({\bf{Estimator side}})}}\\
$\sigma_{k}^{i}=\sqrt{\bmC_{i}\bmP_{k|k}^{i-1}\bmC_{i}'+R_{i}},$\\
Transmit $\bmC_i\hat{\bmx}_{k|k}^{i-1},\sigma_k^i$ back to sensor\\
{\color{black}{({\bf{Sensor side}})\\}}
$z_{k}^{i}=y_{k}^{i}-\bmC_{i}\hat{\bmx}_{k|k}^{i-1},$\\
$\epsilon_{k}^{i}=z_{k}^{i}/\sigma_{k}^{i}.$\\
{Power scheduling:}\\
Define the scheduling variable as\\
$\gamma_{k}^{i}=\left\{\begin{array}{cc}
1,&\text{if}~|\epsilon_{k}^{i}|>\eta_i;\\
0,&\text{otherwise.}
\end{array}
\right.$\\
{Measurement transmission:}\\
If $\gamma_{k}^i=1,$ send $y_{k}^i$ to the estimator with high transmission power $\Delta;$ otherwise, transmit $y_k^i$ with low transmission power $\delta.$\\
If $\gamma_k^i=0,$ then use variable $\beta_k^i$ to represent whether $y_k^i$ is successfully transmitted to the remote estimator. Namely,
$\beta_k^i=1$ indicates $y_k^i$ arrives successfully at the remote estimator and $\beta_k^i=0$ means $y_k^i$ drops during the transmission. \\
{\color{black}{({\bf{Estimator side}})}}\\
{Measurement update:}\\
For $i=1$ to $m,$ let
$\bmK_{k}^{i}=\bmP_{k|k}^{i-1}\bmC_{i}'\left(\bmC_{i}\bmP_{k|k}^{i-1}\bmC_{i}'+\bmR_{i}\right)^{-1},$\\
do\\
$\hat{\bmx}_{k|k}^{i}=\hat{\bmx}_{k|k}^{i-1}+s\left(\gamma_k^i,\beta_k^i\right)\bmK_{k}^{i}
\left(y_{k}^{i}-\bmC_{i}\hat{\bmx}_{k|k}^{i-1}\right),$\\
$\bmP_{k|k}^{i}=\bmP_{k|k}^{i-1}-
t\left(\gamma_k^i,\beta_k^i\right)\bmK_{k}^{i}\bmC_{i}\bmP_{k|k}^{i-1}$\\
where

$s\left(\gamma_k^i,\beta_k^i\right)=\gamma_{k}^i+\left(1-\gamma_k^i\right)\beta_k^i,$\\
$t\left(\gamma_k^i,\beta_k^i\right)=\gamma_{k}^i+\left(1-\gamma_k^i\right)l(\beta_k^i),$\\
$l\left(\beta_{k}^{i}\right)=\beta_{k}^{i}+\left(1-\beta_{k}^{i}\right)\sqrt{\frac{2}{\pi}}
\!\times\!\frac{\eta_i\exp(-\eta_i^{2}/2)}{1-2Q(\eta_i)}.$

End, do $\hat{\bmx}_{k|k}=\hat{\bmx}_{k|k}^{m}$ and $\bmP_{k|k}=\bmP_{k|k}^{m}.$
}
\end{algorithmic}
\end{algorithm}
{\color{black}{{\color{black}{
The mechanism of the proposed innovation-based scheduler will be further elaborated in this part. The power scheduler located at the sensor side will sequentially decide whether to adopt a high or low energy to transmit every single component $y_k^1,y_k^2,\ldots,y_k^m$ to remote estimator. For instance, the scheduler is now in a position to make decision on assigning energy for transmission of $y_k^i$ at time slot $i$ of time instant $k.$ Suppose the estimator has already reliably sent back current state estimate $\hat{\bmx}_{k|k}^{i-1}$ and error covariance $\bmP_{k|k}^{i-1}$ to local sensor side, where, by ``reliably'' we mean the remote estimator can always adopt a high energy to broadcast information to the sensor so no packet drop happens for the transmission from estimator back to sensor side. This also makes sense because usually wireless sensors consume much less energy for receiving one packet than sending one packet \cite{mica}. Thus, after receiving $\hat{\bmx}_{k|k}^{i-1}$ and $\bmP_{k|k}^{i-1},$ the sensor can compute a normalized innovation $\epsilon_k^i$ of current single component $y_k^i$ of measurement vector $\bmy_k.$ Then comparing the normalized innovation with a given threshold $\eta_k^i,$ if greater than the threshold, this component $y_k^i$ will be transmitted to estimator by a high energy; otherwise, a low energy will be used at this time slot $i.$ The estimator will correspondingly leverage different rules to update $\hat{\bmx}_{k|k}^{i-1}, \bmP_{k|k}^{i-1}$ to obtain $\hat{\bmx}_{k|k}^i,\bmP_{k|k}^i$ and then reliably send them back to the local sensor side for the next cycle.
}}
}}


{\color{black}{
\begin{remark}
It should be noticed that the effect of channel medium between the sensor and estimator has not been considered here. As pointed out in \cite{handbook}, most of the existing work in distributed estimation assumed perfect channels between sensors and the fusion center. However, since only one scalar $y_k^i$ is transmitted at every time slot, let us consider transmitting $y_k^i$ to estimator over a channel with channel gain $f_k^i.$ We envision that the channel undergoes slow fading such that the phase of complex channel can be estimated and therefore compensated for at the receiver side, so that $f_k^i$ defines the real-valued envelop of the complex channel gain \cite{auto2009dey}. Also, suppose that the channel gain remains invariant over the time slot to send $y_k^i.$ 
Then the estimator receives a scaled version of $y_k^i$ corrupted with the channel noise which is independent of the measurement noise. For simplicity, let the channel noise $n_k^i$ be zero-mean Gaussian white noise with variance $\sigma_{n_i}^2$ and, let $\{n_k^i\},$ $\{n_k^j\}$ be mutually independent for $i\ne j.$

In light of the round-robin, time-slotted transmission policy and (\ref{measurement}), one further arrives at $\left(y_k^i\right)^{\rm{ out}}=f_k^i\bmC_i\bmx_k+f_k^i\upsilon_k^i+n_k^i.$ Denote 
$\tilde{\bmupsilon}_k^i=f_k^i\bmupsilon_k^i+n_k^i,$ and it follows that
\begin{align}
\left(y_k^i\right)^{\rm{out}}=f_k^i\bmC_i\bmx_k+\tilde{\upsilon}_k^i.\label{channel}
\end{align}
If channel gain $f_k^i$ is constant over time instant $k,$ then by letting $\tilde{\bmC}_k^i=f_k^i\bmC_i,$ this model reduces to (\ref{measurement}). Nonetheless, for faded channel case, Kalman filtering with faded observations was considered in \cite{auto2009dey} and stability analysis was also presented therein. So results in this paper can be generalized to the faded channel case by adopting similar method in \cite{auto2009dey} to tackle $f_k^i$ of fading distribution, which will be explored in the future.
\end{remark}
}
}
\begin{proposition}
\label{proposition}
It is postulated that the conditional distribution of $\bmx_{k}$ given $\mathcal{I}_{k-1}^{i-1}$ is approximately Gaussian, i.e., the probability density function (pdf) $f\left(\bmx_{k}|\mathcal{I}_{k}^{i-1}\right)=\mathcal{N}\big(\bmx_{k};\hat{\bmx}_{k|k}^{i-1},\bmP_{k|k}^{i-1}\big).$
Then $\hat{\bmx}_{k|k}^{i}$ in Algorithm \ref{ps} is a minimum mean squared error estimator.
\end{proposition}

{\color{black}{
\begin{remark}
 Recall further that the pdf $f\left(\bmx_{k}|\mathcal{I}_{k}^{i-1}\right)$ is in general non-Gaussian and therefore, (computationally expensive) numerical integrations and (memory intensive) propagation of the posterior pdf are required for the computation of the exact MMSE estimate \cite{tsp2006ribeiro}. However, based upon customary simplifications adopted in nonlinear filtering \cite{tsp2003gpf} and Kalman filtering with quantized measurements/innovations \cite{tsp2006ribeiro, tsp2013nair, spm2006luo}, the assumption on an approximately Gaussian distribution of the predicted density is made. This assumption can be widely found in the literature; see, for example, \cite{tsp2008msechu, timc2011you, tsp2012you} and references therein for further discussion.
\end{remark}
}
}

\begin{IEEEproof}[Proof of Proposition 1]
Provided that we already have an MMSE estimator $\hat{\bmx}_{k|k}^{i-1},$ that is, $\hat{\bmx}_{k|k}^{i-1}=\mathbb{E}\left[\bmx_k|\mathcal{I}_{k}^{i-1}\right]$ and $\bmP_{k|k}^{i-1}=\mathbb{E}\big[(\bmx_k-\hat{\bmx}_{k|k}^{i-1})
(\bmx_k-\hat{\bmx}_{k|k}^{i-1})'|\mathcal{I}_{k}^{i-1}\big].$ We prove the proposition by conditioning on whether the measurement is received by the estimator. Specifically, when the new measurement $y_k^i$ is present at the estimator side, that is,  the case $\gamma_k^i=1$ or the case $\gamma_k^i=0~\text{and}~\beta_k^i=1,$ one can easily verify that
\begin{align*}
\hat{\bmx}_{k|k}^i&=\mathbb{E}\left[\bmx_k\big|\mathcal{I}_k^{i-1},y_k^i\right]\\
&=\hat{\bmx}_{k|k}^{i-1}+\bmK_k^i\left(y_k^i-\bmC_i\hat{\bmx}_{k|k}^{i-1}\right),
\end{align*}
and likewise,
\begin{align*}
\bmP_{k|k}^i&=\mathbb{E}\left[(\bmx_k-\hat{\bmx}_{k|k}^{i})(\bmx_k-\hat{\bmx}_{k|k}^{i})'\big|
\mathcal{I}_{k}^{i-1},y_k^i\right]\\
&=\bmP_{k|k}^{i-1}-\bmP_{k|k}^{i-1}\bmC_i'\left(\bmC_i\bmP_{k|k}^{i-1}\bmC_i'+R_i\right)^{-1}\bmC_i\bmP_{k|k}^{i-1}.
\end{align*}

When the estimator does not receive the new measurement $y_k^i,$ that is, $\gamma_k^i=0~\text{and}~\beta_k^i=0,$ then it follows that
\begin{align}
\hat{\bmx}_{k|k}^i=&\mathbb{E}\left[\bmx_k\big|\mathcal{I}_{k}^{i-1},
\gamma_k^i=0,\beta_k^i=0\right]\nonumber\\
=&\mathbb{E}\left[\bmx_k\big|\mathcal{I}_{k}^{i-1},
|\epsilon_k^i|\leq\eta_i\right]\nonumber\\
=&\int_{-\eta_i}^{\eta_i}\big(\hat{\bmx}_{k|k}^{i-1}\!+\!\bmK_k^i\sigma_k^i\epsilon\big)
f_{\epsilon_k^i}\left(\epsilon\big|\mathcal{I}_{k}^{i-1},
|\epsilon_k^i|\leq\eta_i\right)d\epsilon\label{pdf}.
\end{align}
{\color{black}{Here, $f_\bbx(x)$ is the pdf of the random variable $\bbx;$ similarly, $f_{\bbx|\bby}(x|y)$ is the pdf of a random variable $\bbx$ conditional on variable $y.$
Given $\mathcal{I}_k^{i-1},$ then $\epsilon_k^i$ follows Gaussian distribution with zero-mean and unit covariance. Thus, the conditional pdf above 
follows directly from conditional probability theory:
\begin{align}
f_{\epsilon_k^i}\left(\epsilon|\mathcal{I}_{k}^{i-1},
|\epsilon_k^i|\leq\eta_i\right)=\left\{\begin{array}{ll}
\frac{f_{\epsilon_k^i}\left(\epsilon|\mathcal{I}_{k}^{i-1}\right)}{\Delta\! P_i},&
{\rm{if}}~|\epsilon_k^i|\leq\eta_i,\\
0,&\rm{otherwise}
\end{array}
\right.\label{how}
\end{align}
where $\vartriangle\!\!P_i\triangleq{\rm{Pr}}\left(|\epsilon_k^i|\leq\eta_i
\big|\mathcal{I}_k^{i-1}\right)=1-2Q(\eta_i).$ Therefore, (\ref{pdf}) becomes
\begin{align}
\hat{\bmx}_{k|k}^{i}=&\int_{-\eta_i}^{\eta_i}
\left(\hat{\bmx}_{k|k}^{i-1}+\bmK_k^i\sigma_k^i\epsilon\right)
\frac{f_{\epsilon_k^i}\left(\epsilon|\mathcal{I}_{k}^{i-1}\right)}
{\vartriangle\!\! P_i}d\epsilon\nonumber\\
=&\hat{\bmx}_{k|k}^{i-1}\int_{-\eta_i}^{\eta_i}
\frac{f_{\epsilon_k^i}\left(\epsilon|\mathcal{I}_{k}^{i-1}\right)}
{\vartriangle\!\! P_i}d{\epsilon}+\nonumber\\
&~\frac{\bmK_k^i\sigma_k^i}{\vartriangle\!\! P_i}\int_{-\eta_i}^{\eta_i}
\epsilon f_{\epsilon_k^i}\left(\epsilon|\mathcal{I}_{k}^{i-1}\right)
d\epsilon\nonumber\\
=&\hat{\bmx}_{k|k}^{i-1}\label{noy}
\end{align}}}where the first integration equals to $1$ and the second becomes $0$ because $f_{\epsilon_k^i}\left(\epsilon|\mathcal{I}_{k}^{i-1}\right)$ is even over the origin-centered symmetric integration interval.

In the sequel, we compute the covariance $\bmP_{k|k}^{i}$ for $\gamma_k^i=0$ and $\beta_k^i=0$ case as follows:
\begin{align}
\bmP_{k|k}^{i}\stackrel{(a)}{\!=\!}&\mathbb{E}\left[\left(\bmx_k-\hat{\bmx}_{k|k}^i\right)\left(\bmx_k-\hat{\bmx}_{k|k}^i\right)'
\Big|\mathcal{I}_k^{i-1},|\epsilon_k^i|\leq \eta_i\right]\nonumber\\
\stackrel{(b)}{=}&\mathbb{E}\left[\left(\bmx_k-\hat{\bmx}_{k|k}^{i-1}\right)
\left(\bmx_k-\hat{\bmx}_{k|k}^{i-1}\right)'
\Big|\mathcal{I}_k^{i-1},|\epsilon_k^i|\leq \eta_i\right]\nonumber\\
\stackrel{(c)}{=}&\mathbb{E}\bigg[\left(\bmx_k-\hat{\bmx}_{k|k}^{i-1}-\bmK_k^i\sigma_k^i\epsilon
+\bmK_k^i\sigma_k^i\epsilon\right)\times\nonumber\\
&\left(\bmx_k\!-\!\hat{\bmx}_{k|k}^{i-1}\!-\!\bmK_k^i\sigma_k^i\epsilon+
\bmK_k^i\sigma_k^i\epsilon\right)'
\Big|\mathcal{I}_k^{i-1},|\epsilon_k^i|\leq \eta_i\bigg]\nonumber\\
\stackrel{(d)}{\!=\!}&\mathbb{E}\bigg[\left(\left(I_n\!-\!\bmK_k^i\bmC_i\right)\big(\bmx_k\!-\!\hat{\bmx}_{k|k}^{i-1}\big)\!{\color{black}{+\!\bmK_k^i\upsilon_k^i}}\!+\!
\bmK_k^i\sigma_k^i\epsilon\right)\nonumber\\
&\!\times\!\left(\left(\bmI_n\!-\!\bmK_k^i\bmC_i\right)\!\big(\bmx_k\!-\!\hat{\bmx}_{k|k}^{i-1}\big)\!{\color{black}{+\!\bmK_k^i\upsilon_k^i}}\!+\!
\bmK_k^i\sigma_k^i\epsilon\right)'\!\Big|\nonumber\\
&\quad\mathcal{I}_k^{i-1},\,|\epsilon_k^i|\leq \eta_i\bigg]\nonumber\\
\stackrel{(e)}{\!=\!}\!&\left(\bmI_n\!-\!\bmK_k^i\bmC_i\right)\bmP_{k|k}^{i\!-1}\left(\bmI_n\!-\!\bmK_k^i\bmC_i\right)'\!{\color{black}{+\!\bmK_k^iR_i(\bmK_k^i)'}}\nonumber\\
&+\left(\sigma_k^i\right)^2 \bmK_k^i
\mathbb{E}\left[\epsilon\big|\mathcal{I}_k^{i-1},
|\epsilon_k^i|\leq \eta_i\right]\left(\bmK_k^i\right)'\label{pkki}
\end{align}
where {\color{black}{(b) follows directly from $\hat{\bmx}_{k|k}^i=\hat{\bmx}_{k|k}^{i-1}$ when the new measurement component $y_k^i$ is not received by the estimator, which has been proved in (\ref{noy}), and (d) is because $\sigma_{k}^{i}\epsilon_{k}^{i}=z_{k}^{i}=y_k^i-\bmC_i\hat{\bmx}_{k|k}^{i-1}=\bmC_i\left(\bmx_k-\hat{\bmx}_{k|k}^{i-1}\right)+\upsilon_k^i$ in Algorithm \ref{ps}, and (e) is because $\upsilon_k^i$ is zero-mean Gaussian noise with covariance $R_i,$ or, $\mathbb{E}\left[\bmK_k^i\upsilon_k^i\left(\bmK_k^i\upsilon_k^i\right)'\right]=\bmK_k^iR_i\left(\bmK_k^i\right)'.$ }} Meanwhile, we have
\begin{align}
&\mathbb{E}\left[\epsilon^2\Big|\mathcal{I}_k^{i-1},
|\epsilon_k^i|\leq \eta_i\right]\nonumber\\
=&{\color{black}{\int_{-\eta_i}^{\eta_i}\epsilon^2f_{\epsilon_k^i}\left(\epsilon|\mathcal{I}_{k}^{i-1},
|\epsilon_k^i|\leq\eta_i\right)d\epsilon}}\nonumber\\
=&\frac{1}{1-2Q(\eta_i)}\int_{-\eta_i}^{\eta_i}\frac{\epsilon^2}
{\sqrt{2\pi}}{\rm{exp}}(-\epsilon^2/2)d\epsilon\nonumber\\
=&1-\sqrt{\frac{2}{\pi}}\times
\frac{\eta_i{\rm{exp}}\left(-\eta_i^2/2\right)}{1-2Q(\eta_i)}.\label{prob}
\end{align}
{\color{black}{Therefore, from (\ref{pkki})-(\ref{prob}) and $\left(\sigma_k^i\right)^2=\bmC_i\bmP_{k|k}^{i-1}\bmC_i'+R_i,~\bmK_k^i=\bmP_{k|k}^i\bmC_i'\left(\bmC_i\bmP_{k|k}^i\bmC_i'+R_i\right)^{-1},$ one arrives at
\begin{align*}
\bmP_{k|k}^{i}\!=&\left[\left(\bmI_n\!-\!\bmK_k^i\bmC_i\right)\bmP_{k|k}^{i-\!1}\left(\bmI_n\!-\!\bmK_k^i\bmC_i\right)'\!+\!\bmK_k^iR_i(\bmK_k^i)'\right]\nonumber\\
&+\left(\sigma_k^i\right)^2 \bmK_k^i\left[1-\sqrt{\frac{2}{\pi}}\frac{\eta_i{\rm{exp}}
\left(-\eta_i^2/2\right)}{1-2Q(\eta_i)}\right] \left(\bmK_k^i\right)'\\
=&\left[\bmP_{k|k}^{i-1}-\bmP_{k|k}^{i-1}\bmC_i'\left(\bmC_i\bmP_{k|k}^i\bmC_i'+R_i\right)^{-1}\bmC_i\bmP_{k|k}^i\right]-\\
&\left(1-\sqrt{\frac{2}{\pi}}\times
\frac{\eta_i{\rm{exp}}\left(-\eta_i^2/2\right)}{1-2Q(\eta_i)}\right)\times\\
&\bmP_{k|k}^i\bmC_i'\left(\bmC_i\bmP_{k|k}^i\bmC_i'+R_i\right)^{-1}\bmC_i\bmP_{k|k}^i\\
=&\bmP_{k|k}^{i-1}-\sqrt{\frac{2}{\pi}}\!\times\!
\frac{\eta_i{\rm{exp}}\left(-\eta_i^2/2\right)}{1-2Q(\eta_i)}
\bmK_k^i\bmC_i\bmP_{k|k}^{i-1}.
\end{align*}

To write the two scenarios discussed above in a more compact form, it follows that
\begin{align}
\bmP_{k|k}^{i}&=\left[\gamma_k^i+\left(1-\gamma_k^i\right)\beta_k^i\right]
\left(\bmP_{k|k}^{i-1}-\bmK_{k}^{i}\bmC_{i}'\bmP_{k|k}^{i-1}\right)+\nonumber\\
&\quad\left(1-\gamma_k^i\right)\left(1-\beta_k^i\right)\Big(\bmP_{k|k}^{i-1}-\sqrt{\frac{2}{\pi}}\!\times\!
\frac{\eta_i{\rm{exp}}\left(-\eta_i^2/2\right)}{1-2Q(\eta_i)}\nonumber\\
&\quad\times\bmK_k^i\bmC_i\bmP_{k|k}^{i-1}\Big)\nonumber\\
&=\left[\gamma_k^i+\left(1-\gamma_k^i\right)\beta_k^i+\left(1-\gamma_k^i\right)\left(1-\beta_k^i\right)\right]\bmP_{k|k}^{i-1}-\nonumber\\
&\quad \bigg[\gamma_k^i+\left(1-\gamma_k^i\right)\beta_k^i-\left(1-\gamma_k^i\right)\left(1-\beta_k^i\right)\sqrt{\frac{2}{\pi}}\times\nonumber\\
&\quad\frac{\eta_i{\rm{exp}}\left(-\eta_i^2/2\right)}{1-2Q(\eta_i)}\bigg]\bmK_k^i\bmC_i\bmP_{k|k}^{i-1}\nonumber\\
&=\bmP_{k|k}^{i-1}-
t\left(\gamma_k^i,\beta_k^i\right)\bmK_{k}^{i}\bmC_{i}\bmP_{k|k}^{i-1},\label{pki}
\end{align}
which completes the proof.}}

\end{IEEEproof}

Given the new filter formulation in Algorithm \ref{ps}, the processes $\{\gamma_{k}^{1}\}_{0}^{\infty},$ $\{\gamma_{k}^{2}\}_{0}^{\infty},
\ldots,\{\gamma_{k}^{m}\}_{0}^{\infty}$ form a sequence of independent and identically distributed (i.i.d.) processes under the Gaussian approximation \cite{tsp2012you} and also, assume the processes $\{\beta_{k}^{1}\}_{0}^{\infty},$ $\{\beta_{k}^{2}\}_{0}^{\infty},
\ldots,\{\beta_{k}^{m}\}_{0}^{\infty}$ are mutually independent Bernoulli i.i.d. processes. Define $N_{k}={\rm{diag}}\left[t(\gamma_k^1),t(\gamma_k^2),\ldots,t(\gamma_k^m)\right].$ For $i=1,2,\ldots,m,$ let
\begin{align}
\mu_i&=\mathbb{E}[\gamma^{i}_{k}]=2Q(\eta_i)\label{mu}\\
\nu_i&=\frac{\eta_i\exp(\eta^2_{i}/2)}
{\sqrt{\pi/2}\left(1-2Q(\eta_i)\right)}\label{nu}
\end{align}
and $\mathbb{E}[\beta_k^i]=\beta;$ in addition,
\begin{align}
\mathbb{E}\left[l(\beta_k^i)\right]&=\mathbb{E}\left[\beta_{k}^{i}+(1-\beta_{k}^{i})
\times\frac{\eta_i\exp(-\eta_i^{2}/2)}{\sqrt{\pi/2}(1-2Q(\eta_i))}\right]\nonumber\\
&=\beta+\left(1-\beta\right)\nu_i=\xi_i,\label{mu}\\
\mathbb{E}\left[t(\gamma_k^i,\beta_k^i)\right]
&=\mathbb{E}\left[\gamma_{k}^{i}+(1-\gamma_{k}^{i})l(\beta_k^i)\right]\nonumber\\
&=\mu_i+\left(1-\mu_i\right)\xi_i=\lambda_i,\label{lambda}
\end{align}
{\color{black}where, in fact, we have $\nu_i=1-1/\sqrt{2\pi}\cdot\int_{-\eta_i}^{\eta_i}{\rm{exp}}(-t^2/2)dt\in[0,1]$ and $\xi_i\in[0,1].$ Moreover, $\nu_i$ is one strictly decreasing function in threshold $\eta_i;$ this makes sense since the greater the threshold is, the less information will be transmitted through high energy. Then, one can easily verify $0\le\lambda_i\le1,$ and therefore, $\lambda_i$ can be somehow physically interpreted as the normalized averaged information received by remote estimator resulting from the power scheduling and networked effect on transmitting $y_k^i$ (and $1\!-\lambda_i$ quantifies the corresponding averaged information loss rate). All $\lambda_i$s together will governor the mean square stability of estimation error covariance matrix, which will be investigated in the ensuing section. Therefore, we will refer to $\lambda_i$ hereafter other than the specific parameters $\eta_i,\beta.$}


\section{Statistical Properties and Sufficient, Necessary Convergence Conditions}
\label{section c}
{\color{black}{In this section, the convergence conditions for the expected estimation error covariance will be provided by discussing properties of a constructed function. Denote $\{\gamma_k^i\}:=\big\{\{\gamma_{k}^{1}\}_{0}^{\infty},$ $\{\gamma_{k}^{2}\}_{0}^{\infty},
\ldots,\{\gamma_{k}^{m}\}_{0}^{\infty}\big\}$ and $\{\beta_k^i\}:=\big\{\{\beta_{k}^{1}\}_{0}^{\infty},$ $\{\beta_{k}^{2}\}_{0}^{\infty},
\ldots,\{\beta_{k}^{m}\}_{0}^{\infty}\big\}.$ Since they are inherently stochastic and cannot be determined offline, therefore, only statistical properties can be derived. Before delving into main results, some preliminaries will be given in the following.}}

Let $\mathbb{S}_{+}^{n}=\{\bmS\in \mathbb{R}^{n\times n}|\bmS\ge {\bm0}\}.$  Define the function $\bmh:\mathbb{S}_{+}^{n}\to\mathbb{S}_{+}^{n}$ and the function $\bmg_{\lambda_{i}}:\mathbb{S}_{+}^{n}\to\mathbb{S}_{+}^{n}$ as follows:
\begin{align}
\bmh(\bmX)&\triangleq \bmA\bmX\bmA'+\bmQ\\
\bmg_{\lambda_{i}}(\bmX)&\triangleq\bmX-\lambda_{i}\bmX\bmC_{i}'\left(\bmC_{i}\bmX\bmC_{i}'+R_{i}\right)^{-1}\bmC_{i}\bmX\\
{\color{black}\bmg_{\lambda_i}}\!\circ\bm h(\bmX)&\triangleq {\color{black}\bmg_{\lambda_i}}\!(\bmh(\bmX))
\end{align}
and here denote the notation $\circ$ by the function composite.
Therefore, the covariance update in the sequential Kalman filter formulation in Algorithm \ref{ps} becomes
\begin{align*}
\bmP_{k|k-1}&=\bmh(\bmP_{k-1|k-1})\\
\bmP_{k|k}^{1}&=\bmg_{\lambda_{1}}(\bmP_{k|k}^{0})=\bmg_{\lambda_{1}}(\bmP_{k|k-1})\\
\bmP_{k|k}^{i}&=\bmg_{\lambda_{i}}(\bmP_{k|k}^{i-1}),i=2,3,\ldots,m-1\\
{\color{black}\bmP_{k|k}}\!=\bmP_{k|k}^{m}&=\bmg_{\lambda_{m}}(\bmP_{k|k}^{m-1}).
\end{align*}
Let
\begin{align}
\bmP_{k|k-1}&=\bmh(\bmP_{k-1|k-1})\label{h function}\\
\bmP_{k|k}&=\mathcal{M}_{m}(\bmP_{k|k-1})\triangleq \bmg_{\lambda_{m}} \bmg_{\lambda_{m-1}}\ldots \bmg_{\lambda_{1}}(\bmP_{k|k-1}).\label{covariance}
\end{align}
Denote the function $\bm{\varphi}:\mathbb{S}^{n}\to\mathbb{S}^{n}$ by the transformation from $\bmP_{k-1|k-1}$ to $\bmP_{k|k},$ namely,
\begin{align}
\bmP_{k|k}=\bm{\varphi}(\bmP_{k-1|k-1})\triangleq \mathcal{\bmM}_{m}h(\bmP_{k-1|k-1}).
\end{align}

In order to analyze the convergence of the estimation error covariance matrix, we then define the modified algebraic Riccati equation (MARE) in the following way:
\begin{align}
\bm{\varphi}(\bmP_{k})&=\bmg_{\lambda_{m}}\bmg_{\lambda_{m-1}}\ldots \bmg_{\lambda_{1}}\bmh(\bmP_{k})\label{varphi function}
\end{align}
where we used the simplified notation $\bmP_{k}=\bmP_{k|k},k\ge 0.$ {\color{black}{Meanwhile, as explained, the covariance matrices $\{\bmP_{k}\}_0^{\infty}$ depend nonlinearly on the specific realization of the stochastic processes $\{\gamma_k^i\}$ and $\{\beta_k^i\},$ so the sequential Kalman filter is inherently stochastic and cannot be determined offline. Then, only statistical properties with respect to the covariance matrices of the proposed sequential Kalman filter can therefore be established.}}
\begin{remark}
\label{remark mare}
It is noted in passing that the modified algebraic Riccati equation defined in (\ref{varphi function}) is a more generalized form than the original MARE specified for Kalman filtering with only one or two lossy channels in \cite{tac2004sinopoli} and \cite{cdc2004goldsmith}, respectively, where the analysis might be much easier than that of (\ref{varphi function}). Moreover, since the MARE in (\ref{varphi function}) is sequentially composited by $m$ original MAREs with different parameters, namely, $\lambda_1,\lambda_2,\ldots,\lambda_m,$ then the MARE in (\ref{varphi function}) is also quite different from the MARE discussed in \cite{tac2012sinopoli} defined for Kalman filtering for multiple-input multiple-output systems with control signals and sensored measurements transmitting across multiple TCP-like erasure channels. Accurately speaking, the MARE in \cite{tac2012sinopoli} follows directly from that in \cite{tac2004sinopoli} by replacing the observation matrix $\gamma_{k}\bmC$ with $\text{diag}\{\gamma_{k}^1,\gamma_k^2,\ldots,\gamma_k^m\}\bmC.$ Therefore, for the sake of completeness on stability theory of Kalman filtering with intermittent observations, the investigation on properties of the MARE in (\ref{varphi function}) in the following sections is also of great implications, which significantly contributes to the derivation of sufficient conditions for stability of sequential Kalman filtering with scheduled measurements in \cite{tsp2012you}.
\end{remark}

 The following lemma on the properties of the auxiliary function $\bm{\psi}_{\lambda_{i}}$ is presented before we will formally study the convergence properties of the MARE in (\ref{varphi function}).
\begin{lemma}[\cite{tac2004sinopoli}]%
\label{psi lemma}
Let the function $\bm{\psi}_{\lambda_{i}}$ be
\begin{align}%
\label{psi function}
\bm{\psi}_{\lambda_{i}}(\bm\bmL_{i},\bmX)=(1-\lambda_{i})\bmX+&\lambda_{i}\left(\bm\bmE_{i}\bmX\bm\bmE_{i}'+\bmL_{i}R_{i}\bmL_{i}'\right),\nonumber\\
&\qquad\qquad\quad i=1,\ldots,m
\end{align}
where $\bm\bmE_{i}=\bmI_n+\bmL_{i}\bmC_{i}, R_{i}>0,\bmX,\bmY,\bmZ\in \mathbb{S}_{+}^{n}.$ 
Then the following facts hold:
\begin{enumerate}
\item With given $\bmL_{i}^{\bmX}=-\bmX\bmC_{i}'(\bmC_{i}\bmX\bmC_{i}'+R_{i})^{-1},$ $\bmg_{\lambda_{i}}(\bmX)=\bm{\psi}_{\lambda_{i}}(\bmL_{i}^{\bmX},\bmX)$\label{fact11}
\item $\bmg_{\lambda_{i}}(\bmX)=\min_{\bmL_{i}}\bm{\psi}(\bmL_{i},\bmX)\leq\bm{\psi}(\bmL_{i},\bmX),\forall \bmL_{i}$\label{fact12}
\item If $\bmX\leq \bmY,$ then $\bmg_{\lambda_{i}}(\bmX)\leq \bmg_{\lambda_{i}}(\bmY)$\label{fact13}
\item If $\lambda_{i}\ge\lambda_{j},$ then $\bmg_{\lambda_{i}}(\bmX)\le \bmg_{\lambda_{j}}(\bmX)$\label{fact14}
\item If $\tau\in[0,1],$ then $\bmg_{\lambda_{i}}(\tau \bmX+(1-\tau)\bmY)\geq \tau \bmg_{\lambda_{i}}(\bmX)+(1-\tau)\bmg_{\lambda_{i}}(\bmY)$\label{fact15}.
\end{enumerate}
\end{lemma}
\begin{IEEEproof}
The proofs for these statements are analogous to those of Lemma 1 in \cite{tac2004sinopoli} with some appropriate notation adaptations.
\end{IEEEproof}

Notice, that the relationship between the function $\bmg_{\lambda_{i}}$ and the function $\bm{\psi}_{\lambda_i}$ has been built, and now, in order to investigate the convergence properties of the MARE in (\ref{varphi function}), the relationship between the composite function $\bmg_{\lambda_m}\bmg_{\lambda_{m-1}}\ldots \bmg_{\lambda_1}$ and the introduced auxiliary function $\bm{\psi}_{\lambda_m}\bm{\psi}_{\lambda_{m-1}}\ldots\bm{\psi}_{\lambda_1}$ will be constructed in the following way.

 According to (\ref{psi function}), observe that the function $\bm{\psi}_{\lambda_{i}}(\bmL_{i},\bmX)$ is a function with respect to two matrix variables $\bmL_i,\bmX$. With a slight abuse of notation, denote $\bm{\psi}_{\lambda_{j+1}}\bm{\psi}_{\lambda_{j}}(\bmL_{j+1},\bmL_{j},\bmX)$ by the composite function $\bm{\psi}_{\lambda_{j+1}}\left(\bmL_{j+1},\bm{\psi}_{\lambda_{j}}\left(\bmL_{j},\bmX\right)\right)$ with respect to the second variable $\bmX,$ where $j=1,2,\ldots,m-1.$

In the sequel, 
$\mathcal{T}_{s}\left(\bmL_{1},\bmL_{2},\ldots,\bmL_{s},X\right),s=1,2,\ldots, m$ can be derived as follows.
Let us define $\mathcal{T}_{s}\left(\bmL_{1},\bmL_{2},\ldots,\bmL_{s},\bmX\right)=\bm{\psi}_{\lambda_{s}}\bm{\psi}_{\lambda_{s-1}}\ldots\bm{\psi}_{\lambda_{1}}(\bmL_{1},\bmL_{2},\ldots,\bmL_{s},\bmX),$ $s=1,2,\ldots,m.$ Then,
\begin{align}%
\mathcal{T}_{s}=&\bm{\psi}_{\lambda_{s}}(\bmL_{s},\mathcal{T}_{s-1})\nonumber\\
=&(1-\lambda_{s})\mathcal{T}_{s-1}+\lambda_{s}
\left(\bm\bmE_{s}\mathcal{T}_{s-1}\bm\bmE_{s}'+\bmL_{s}R_{s}\bmL_{s}'\right)\nonumber\\
=&\sum_{j=1}^{s-1}\prod_{i=j+1}^{s}(1-\lambda_{i})\lambda_{j}\left(\bm\bmE_{j}\mathcal{T}_{j-1}\bm\bmE_{j}'+\bmL_{j}R_{j}\bmL_{j}'\right)\nonumber\\
+&\prod_{i=1}^{s}(1-\lambda_{i})\bmX+\lambda_{s}\left(\bm\bmE_{s}\mathcal{T}_{s-1}\bm\bmE_{s}'+\bmL_{s}R_{s}\bmL_{s}'\right)\nonumber\\
=&\sum_{j=0}^{s-1}\prod_{i=j+1}^{s}(1-\lambda_{i})\lambda_{j}\left(\bm\bmE_{j}\mathcal{T}_{j-1}\bm\bmE_{j}'+\bmL_{j}R_{j}\bmL_{j}'\right)\nonumber\\
&+\lambda_{s}\left(\bm\bmE_{s}\mathcal{T}_{s-1}\bm\bmE_{s}'+\bmL_{s}R_{s}\bmL_{s}'\right)\label{ts}
\end{align}
where, to make the expression more concrete, we defined $\lambda_{0}=1$ and $\bm\bmE_{0}=\bmI_n,R_{0}=0,\mathcal{T}_{-1}=\bmX, \mathcal{T}_{0}(\bmX)=\bmX.$

For the sake of brevity, denote
\begin{align}%
&\eta_{j,s}^{2}=\prod_{i=j+1}^{s}(1-\lambda_{i})\lambda_{j},j=0,1,\ldots,s-1,\nonumber\\
&\eta_{s,s}^{2}=\lambda_{s}.\label{etajs2}
\end{align}
More importantly, it is easy to exploit the fact that the sum of $s+1$ coefficients $\eta_{j,s}^{2},j=0,1,\ldots,s$ is identically 1, i.e.,
\begin{align*}
\sum_{j=0}^{s}\eta_{j,s}^{2}&=\sum_{j=0}^{s-1}
\Big[\prod_{i=j+1}^{s}(1-\lambda_{i})\lambda_{j}\Big]+{\color{black}{\eta^2_{s,s}}}\\
&=(1-\lambda_{s})+\lambda_{s}=1.
\end{align*}
Alternatively, (\ref{ts}) can be given by
\begin{align}%
&\mathcal{T}_{-1}=\bmX,\mathcal{T}_{0}=\bmX,\nonumber\\
&\mathcal{T}_{s}=\sum_{j=0}^{s}\eta_{j,s}^{2}\left(\bm\bmE_{j}
\mathcal{T}_{j-1}\bm\bmE_{j}'\!+\!\bmL_{j}R_{j}\bmL_{j}'\right),s=1,2,\ldots,m\label{eqts}.
\end{align}
Therefore,
\begin{align}%
\mathcal{T}_{m}=\sum_{j=0}^{m}\eta_{j,m}^{2}
\left(\bm\bmE_{j}\mathcal{T}_{j-1}\bm\bmE_{j}'+\bmL_{j}R_{j}\bmL_{j}'\right)\label{tm function}
\end{align}
where $\eta_{0,0}^{2}=1, \bm\bmE_{0}=\bmI_n,R_{0}=0$ and $\bmX\ge \bm0,$ and $\mathcal{T}_{j}$ is defined in 
(\ref{eqts}) with $\eta_{j,m}$ given by 
(\ref{etajs2}).


\begin{remark}
Note that the auxiliary function $\mathcal{T}_{m}$ defined by (\ref{tm function}) is of similar form to that in \cite{tac2012sinopoli}, \cite{arxiv2009sinopoli}. To be more specific, the latter auxiliary function is referred as follows:
\begin{align}
\bm{\phi}(\bmK_1,\bmK_2,\ldots,\bmK_{2^{m}},\bmP)=\sum_{i=0}^{2^{m}}\widebar{\gamma}_{i}\left(\bmF_{i}\bmY\bmF_{i}'+\bmV_{i}\right)\label{lqg}
\end{align}
with $\bmF_i=\bmA+\bmK_i\bmC_i, \bmV_i=\bmK_iR_i\bmK_i'+\bmQ$ and constants $\widebar{\gamma}_i.$
One can easily observe that there are $2^{m}$ terms in (\ref{lqg}), and the computation burden will become catastrophic when the dimension $m$ of the measurement vector tends to be very large. Meanwhile, only $\frac{m(m+1)}{2}$ terms will be needed for the sequential Kalman filter in this paper, which may significantly reduce the computation burden and is therefore of great importance.
\end{remark}

We are now in a position to establish some properties of the function $\mathcal{\bmT}_{m}(\bmL_{1},\bmL_{2},\ldots,\bmL_{m},\bmX)$ in form of lemmas in the following.
\begin{lemma}%
\label{m lemma}
Consider the function $\mathcal{\bmT}_{m}(\bmL_{1},\bmL_{2},\ldots,\bmL_{m},\bmX)$ as stated by (\ref{tm function}) with $\bm\bmE_{j}=\bmI_n+\bmL_{j}\bmC_{j}.$ Assume $\bmX,\bmY,\bmZ\in \mathbb{S}_{+}^{n}.$ Then, the following facts hold:
\begin{enumerate}
\item With given $\bmL_{j}^{X}=-\mathcal{T}_{j-1}^{\bmX}\bmC_{j}'
    \left(\bmC_{j}\mathcal{\bmT}_{j-1}^{\bmX}\bmC_{j}'+R_{j}\right)^{-1},
    $ $j=1,2,\ldots,m,
    \mathcal{\bmM}_{m}(\bmX)=\mathcal{\bmT}_{m}\left(\bmL_{1}^{\bmX},
    \bmL_{2}^{\bmX},
    \ldots,\bmL_{m}^{\bmX},\bmX\right),$
    where $\mathcal{T}_{j-1}^{\bmX}=\mathcal{T}_{j-1}(\bmL_1^{\bmX},\bmL_2^{\bmX},\ldots,\bmL_{j-1}^{\bmX},\bmX)$\label{fact 1}
\item $\mathcal{\bmM}_{m}(\bmX)=\min_{\bmL_{1},\bmL_{2},\ldots,\bmL_{m}}\mathcal{T}_{m}
    \left(\bmL_{1},\bmL_{2},\ldots,\bmL_{m},\bmX\right)
    \leq\mathcal{T}_{m}\left(\bmL_{1},\bmL_{2},\ldots,\bmL_{m},\bmX\right),\forall \bmL_{1},\bmL_{2},\ldots,\bmL_{m}\in\mathbb{R}^{n\times 1}$ \label{fact 2}
\item If $\bmX\leq \bmY,$ then $\mathcal{\bmM}_{m}(\bmX)\leq \mathcal{\bmM}_{m}\left(\bmY\right)$ \label{fact 3}
\item If $\tau\in[0,1],$ then $\mathcal{\bmM}_{m}(\tau \bmX+(1-\tau)\bmY)\geq \tau \mathcal{\bmM}_{m}(\bmX)+(1-\tau)\mathcal{\bmM}_{m}(\bmY)$\label{fact 5}
\item $\mathcal{M}_{m}(\bmX)\ge \prod_{j=1}^{m}(1-\lambda_{j})\bmX$\label{fact 6}
\item For a random variable $\bmX,$ $\prod_{j=1}^{m}(1-\lambda_{j}){\mathbb{E}}[\bmX]\leq {\mathbb{E}}[\mathcal{\bmM}_{m}(\bmX)]\leq \mathcal{M}_{m}({\mathbb{E}}[\bmX]).$\label{fact 7}
\end{enumerate}
\end{lemma}
\begin{IEEEproof}

\begin{enumerate}
\item\label{proof1} Fact \ref{fact 1}) together with Fact \ref{fact 2}) is equivalent to showing the minimizer and the minimum value of matrix-valued function $\mathcal{\bmT}_s,\forall s=1,2,\ldots,m,$ with respect to multiple vector-valued variables $\bmL_1,\bmL_2,\ldots,\bmL_s\in\mathbb{R}^{n\times 1}.$ For convenience of notation, denote $\bm{\ell}_s=\left(\bmL_1,\bmL_2,\ldots,\bmL_s\right).$ We first make extensive use of differential of general matrix-valued function $\bmF$ with respect to a matrix argument $\bmX;$ see, for instance, \cite{tif2009payaro}. 
   \newline
    \begin{definition}
    Let $\bmF$ be a differentiable $m\times n$ real matrix function of a $p\times q$ matrix of real variables $\bmX.$ The {\emph{Jacobian matrix}} of $\bmF$ at $\bmX$ is given by the $mn\times pq$ matrix
    \begin{align*}
    {\rm{D_{\bmX}}}\bmF(\bmX)=\frac{\partial\, {\rm{vec}}\bmF(\bmX)}{\partial\left({\rm{vec}}\bmX\right)'}
    \end{align*}
    \end{definition}
%

    Then by vectorizing the differential ${\rm{d}}\mathcal{\bmT}_s,$ it gives that:
    \begin{align*}
    {\rm{d}}{\rm{vec}}\mathcal{\bmT}_s\!=\bmJ_{1,s}\,{\rm{d}}{\rm{vec}}\bmL_1+\!
    \bmJ_{2,s}\,{\rm{d}}{\rm{vec}}\bmL_2\!+\ldots\!+\bmJ_{s,s}\,{\rm{d}}{\rm{vec}}\bmL_s
    \end{align*}
    where the Jacobian matrix of matrix $\mathcal{\bmT}_{s}$ with respect to matrix variable $\bmL_{i}$ is defined as $\bmJ_{i,s}=\bmJ_{i,s}(\bmL_1,$ $\bmL_2,\ldots,\bmL_{s})
    ={\rm{\bmD_{\bmL_{i}}}}\mathcal{\bmT}_{s},
    i=1,2,\ldots,s.$ To make the results more concrete, let us define:
    \begin{align*}
    \bmG_{j,j}\triangleq\, &\bmI_n\otimes \bmI_n=\bmI_{n^{2}}\\
    \bmG_{j,t}\triangleq\, &\eta_{j,t}^2\bmI_{n^{2}}+
    \sum_{i=j+1}^{t}\eta_{i,t}^2\left(\bm\bmE_i\otimes \bm\bmE_i\right)\bmG_{j,i-1},\\
    &\qquad\qquad\qquad\qquad t=j+1,j+2,\ldots,s.
    \end{align*}
    Therefore, after complicated and tedious matrix computations, the Jacobian matrices can be obtained as follows:
    \begin{align*}
    \bmJ_{j,s}=&\Big(\eta_{j,s}^2 \bmI_{n^{2}}+\sum_{k=j+1}^s \eta_{k,s}^2\left(\bm\bmE_k\otimes \bm\bmE_k\right)\bmG_{j,k}\Big)\\
    &\times\Big(\left(\bm\bmE_j\mathcal{\bmT}_{j-1}\bmC_j'+\bmL_j\bmR_j\right)\otimes \bmI_n\\
    &+\bmI_{n}\otimes\left(\bm\bmE_j\mathcal{\bmT}_{j-1}\bmC_j'+\bmL_j\bmR_j\right)\Big),\\
    &\qquad\qquad\qquad\qquad\qquad j=1,2,\ldots,s-1.\\
    \bmJ_{s,s}
    =&\eta_{s,s}^2 \Big(\left(\bm\bmE_s\mathcal{\bmT}_{s-1}\bmC_{s}'
    +\bmL_s \bmR_s\right)\otimes \bmI_n\\
    &+\bmI_n\otimes\left(\bm\bmE_s\mathcal{\bmT}_{s-1}\bmC_{s}'
    +\bmL_s \bmR_s\right)\Big)
    \end{align*}
    where intentionally, $\eta_{s,s}^2$ was not replaced by $1$ for the compactness of the structure of $\bmJ_{j,s}.$

    By solving $\bmJ_{j,s}={\bm0},j=1,2,\ldots,s-1,$ it follows straightforwardly that
    \begin{align*}
    &\bm\bmE_j\mathcal{\bmT}_{j-1}\bmC_{j}'+\bmL_jR_j={\bm0}\\
    \Longrightarrow &\bmL_{j,s}^*=-\mathcal{\bmT}_{j-1}\bmC_j'\left(\bmC_j\mathcal{\bmT}_{j-1}\bmC_j'+R_j\right)^{-1}
    \triangleq \bmL_{j}^{\bmX}
    \end{align*}
    where $\mathcal{\bmT}_{j-1}=\mathcal{\bmT}_{j-1}\Big(\bmL_{1}^{\bmX},
    \bmL_{2}^{\bmX},\ldots,\bmL_{j-1}^{\bmX}\Big)
    =\mathcal{\bmT}_{j-1}^\bmX.$

    Then similarly, by solving $\bmJ_{s,s}=\bm0,$ it gives that
    \begin{align*}
    &\bm\bmE_s\mathcal{\bmT}_{s-1}\bmC_{i}'+\bmL_s\bmR_s={\bm0}\\
    \Longrightarrow &\bmL_{s,s}^*=-\mathcal{\bmT}_{s-1}\bmC_s'\left(\bmC_s\mathcal{\bmT}_{s-1}\bmC_s'+R_s\right)^{-1}
    \triangleq \bmL_{s}^{\bmX}
    \end{align*}
    where $\mathcal{\bmT}_{s-1}=\mathcal{\bmT}_{s-1}\Big(\bmL_{1}^{\bmX},
    \bmL_{2}^{\bmX},\ldots,\bmL_{s-1}^{\bmX}\Big)
    =\mathcal{\bmT}_{s-1}^\bmX.$ It should be clearly noticed that $\bmL_{j,s}^*=\bmL_{j,t}^*,$ $\forall t\ge s,$ and then plugging $\bmL_1^\bmX,\bmL_2^\bmX,\ldots,\bmL_m^\bmX$ into
    (\ref{tm function}) verifies that $\mathcal{\bmM}_m(\bmX)=\mathcal{\bmT}_m(\bmL_1^\bmX,\bmL_2^\bmX,\ldots,\bmL_m^\bmX).$

    \item We show this fact by mathematical induction. When $m=1,$ one can easily verify that $\bmL_1^\bmX$ minimizes $\mathcal{\bmT}_{1}(\bmL_1,\bmX).$
        Suppose now that it holds for $m=k;$ that is, the point $(\bmL_1^\bmX,\bmL_2^\bmX,\ldots,\bmL_k^\bmX)$ minimizes $\mathcal{\bmT}_k(\bmL_1,\bmL_2,\ldots,\bmL_k,\bmX).$ Then for $m=k+1,$
    \begin{align*}
    \mathcal{\bmT}_{k+1}=&(1-\lambda_{k+1})\mathcal{\bmT}_{k}\\
    &+\lambda_{k+1}\left(\bm\bmE_{k+1}\mathcal{\bmT}_{k}\bm\bmE_{k+1}'
    +\bmL_{k+1}R_{k+1}\bmL_{k+1}'\right]
    \end{align*}
    and
    \begin{align*}
    {\rm{D}}_{\mathcal{\bmT}_{k}}\mathcal{\bmT}_{k+1}=&(1-\lambda_{k+1})(\bmI_n\otimes \bmI_n)+\lambda_{k+1} (\bm\bmE_{k+1}\otimes \bm\bmE_{k+1})\\
>&{\bm0}
    \end{align*}
    so one necessary condition for some point $\big(\bmL_1^*,\bmL_2^*,\ldots,\bmL_k^*,\bmL_{k+1}^*\big)$ minimizing $\mathcal{\bmT}_{k+1}$ is that the point should also minimize $\mathcal{\bmT}_{k},$
    or, $\big(\bmL_1^*,\bmL_2^*,\ldots,\bmL_k^*\big)$ minimizes $\mathcal{T}_{k}.$ Therefore, $\big(\bmL_1^*,\bmL_2^*,\ldots,\bmL_k^*\big)=\big(\bmL_1^\bmX,\bmL_2^\bmX,\ldots,\bmL_k^\bmX\big),$ or, $\mathcal{\bmT}_k=\mathcal{\bmT}_{k}^\bmX$ when minimizing $\mathcal{\bmT}_{k+1}.$ Given that $\bmL_{k+1}$ is independent of $\mathcal{\bmT}_{k}$ and, $\mathcal{\bmT}_{k}^\bmX>{\bm0},R_{k+1}>0,$ and meanwhile, $\mathcal{\bmT}_{k+1}$ is quadratic and convex in the variable $\bmL_{k+1},$ and therefore, the minimizer for $\mathcal{\bmT}_{k+1}$ can be found by letting
    \begin{align*}
     {\rm{D}}_{\bmL_{k+1}}\mathcal{\bmT}_{k+1}=&\lambda_{k+1}\Big[\left(\bm\bmE_{k+1}\mathcal{\bmT}_k^X\bmC_{k+1}'\!+\!\bmL_{k+1}R_{k+1}\right)\otimes \bmI_n\\
     &+\bmI_n\otimes \left(\bm\bmE_{k+1}\mathcal{\bmT}_k^X\bmC_{k+1}'+\bmL_{k+1}R_{k+1}\right)\Big]\\
     =&{\bm0}
    \end{align*}
    which leads to the unique solution $\bmL_{k+1}^\bmX=-\mathcal{\bmT}_{k}^\bmX\bmC_{k+1}'\left(\bmC_{k+1}\mathcal{\bmT}_{k}^\bmX\bmC_{k+1}'+R_{k+1}\right)^{-1}.$
    Therefore, the point $\left(\bmL_1^\bmX,\bmL_2^\bmX,\ldots,\bmL_k^\bmX,\bmL_{k+1}^\bmX\right)$ minimizes $\mathcal{\bmT}_{k+1}.$ This completes the proof.

\item Observe that the function $\mathcal{\bmT}_{m}$ is affine in the variable $\bmX.$ Let $\bmX\leq \bmY,$ and it yields that
    \begin{align*}
    \mathcal{\bmM}_{m}\left(\bmX\right)
    =&\mathcal{\bmT}_{m}\left(L_{1}^{\bmX},\bmL_2^\bmX,
    \ldots,\bmL_{m}^{\bmX},\bmX\right)\\
    \overset{\text{(a)}}{\leq} &\mathcal{\bmT}_{m}\left(\bmL_{1}^{\bmY},\bmL_2^\bmY,
    \ldots,\bmL_{m}^{\bmY},\bmX\right)\\
    \overset{\text{(b)}}{\leq} &\mathcal{\bmT}_{m}\left(\bmL_{1}^{\bmY},\bmL_2^\bmY,
    \ldots,\bmL_{m}^{\bmY},\bmY\right)\\
    \overset{\text{(c)}}{=}&\mathcal{\bmM}_{m}\left(\bmY\right)
    \end{align*}
    where (a) is because $\mathcal{\bmL}_m^{\bmX}$ minimizes the function $\mathcal{\bmT}_{m}$ with respect to variables $\bmL_1,\bmL_2,\ldots,\bmL_m,$ then for any $\bm{\ell}_m\neq \bm{\ell}_m^{\bmX},$ say, $\bm{\ell}_m=\bm{\ell}_m^\bmY,$ that is, (a) holds true. (b) is due to $\mathcal{\bmT}_{m}$ is affine in the variable $\bmX$ and (c) follows straightforwardly from Fact \ref{fact 2}) above.


\item Let $\bmZ=\tau \bmX+(1-\tau)\bmY,$ where $\tau\in[0,1].$ Notice that
    \begin{align*}
    \mathcal{\bmM}_{1}(\bmZ)&=\mathcal{\bmT}_{1}\big(\bmL_1^{\bmZ},\bmZ\big)\\
    =&\eta_{0,1}^2 \bmZ+\eta_{1,1}^2 \Big[\big(\bmI_n+\bmL_1^\bmZ \bmC_1\big)\bmZ\big(\bmI_n+\bmL_1^\bmZ \bmC_1\big)'\\
    &+\tau \bmL_1^\bmZ R_1\big(\bmL_1^\bmZ\big)'+(1-\tau)\bmL_1^\bmZ R_1(\bmL_1^\bmZ)'\Big]\\
    =&\tau\Big[\eta_{0,1}^2 \bmX+\eta_{1,1}^2\big((\bmI_n+\bmL_1^\bmZ \bmC_1)\bmX(\bmI_n+\bmL_1^\bmZ\bmC_1)'\\
    &+\bmL_1^\bmZ R_1(\bmL_1^\bmZ)'\big)\Big]+(1-\tau)\Big[\eta_{0,1}^2 \bmY
    +\eta_{1,1}^2\\
    &\times\!\Big((\bmI_n\!+\bmL_1^\bmZ \bmC_1)\bmY(\bmI_n+\bmL_1^\bmZ\bmC_1)'\nonumber\\
    &+\bmL_1^\bmZ R_1(\bmL_1^\bmZ)'\Big)\Big]\\
    =&\tau \mathcal{\bmT}_{1}(\bmL_1^\bmZ,\bmX)+(1-\tau)\mathcal{\bmT}_{1}(\bmL_1^\bmZ,\bmY)\\
    \ge & \tau \mathcal{\bmT}_{1}(\bmL_1^\bmX,\bmX)+(1-\tau)\mathcal{\bmT}_{1}(\bmL_1^\bmY,\bmY)\\
    =&\tau\mathcal{\bmM}_{1}(\bmX)+(1-\tau)\mathcal{\bmM}_{1}(\bmY).
    \end{align*}
    Assume that $\mathcal{\bmM}_{s}(\bmZ)\ge\tau \mathcal{\bmM}_s(\bmX)+(1-\tau)\mathcal{\bmM}_s(\bmY).$ Then, we have
    \begin{align*}
    \mathcal{\bmM}_{s+1}&(\bmZ)=\mathcal{\bmT}_{s+1}(\bmL_1^\bmZ,\bmL_2^\bmZ,\ldots,\bmL_{s+1}^\bmZ,\bmZ)\\
    =&(1-\lambda_{s+1})\mathcal{\bmM}_{s}(\bmZ)+\lambda_{s+1}\Big[(\bmI_n+\bmL_{s+1}^\bmZ \bmC_{s+1})\\
    &\times\mathcal{\bmM}_{s}(\bmZ)(\bmI_n+\bmL_{s+1}^\bmZ \bmC_{s+1})'\\
    &+\bmL_{s+1}^\bmZ R_{s+1}(\bmL_{s+1}^\bmZ)'\Big]\\
    \ge&(1-\lambda_{s+1})\Big[\tau \mathcal{\bmM}_s(\bmX)+(1-\tau)\mathcal{\bmM}_s(\bmY)\Big]\\
    &+\lambda_{s+1}\Big[(\bmI_n+\bmL_{s+1}^\bmZ {\bmC}_{s+1})\big(\mathcal{\bmM}_s(\bmX)\\
    &+(1-\tau)\mathcal{\bmM}_s(\bmY)\big)(\bmI_n+\bmL_{s+1}^\bmZ \bmC_{s+1})'\\
    &+(\tau+1-\tau)\bmL_{s+1}^\bmZ R_{s+1}(\bmL_{s+1}^\bmZ)'\Big]\\
    =&\tau\mathcal{\bmT}_{s+1}(\bmL_1^\bmX,\bmL_2^\bmX,\ldots,\bmL_{s}^\bmX,\bmL_{s+1}^\bmZ,\bmX)\\
    &+(1-\tau)\mathcal{\bmT}_{s+1}(\bmL_1^\bmY,L_2^\bmY,\ldots,\bmL_{s}^\bmY,\bmL_{s+1}^\bmZ,\bmY)\\
    \ge&\tau\mathcal{\bmT}_{s+1}(\bmL_1^\bmX,\bmL_2^\bmX,\ldots,\bmL_{s}^\bmX,\bmL_{s+1}^\bmX,\bmX)\\
    &+(1-\tau)\mathcal{\bmT}_{s+1}(\bmL_1^\bmY,\bmL_2^\bmY,\ldots,\bmL_{s}^\bmY,\bmL_{s+1}^\bmY,\bmY)\\
    \ge&\tau\mathcal{\bmM}_{s+1}(\bmX)+(1-\tau)\mathcal{\bmM}_{s+1}(\bmY).
    \end{align*}
    Therefore, the fact holds true.
\item Note that
    \begin{align*}
    \mathcal{\bmM}_{m}(\bmX)=&\mathcal{\bmT}_{m}\left(\bmL_1^\bmX,\bmL_2^\bmX,\ldots,\bmL_m^\bmX,\bmX\right)\\
    =&\eta_{0,m}^2\left(\bm\bmE_0\bmX\bm\bmE_0'+\bmL_0R_0\bmL_0\right)\\
+&\sum_{j=1}^{m}\eta_{j,m}^{2}
    \left[\bm\bmE_{j}^\bmX\mathcal{\bmT}_{j-1}(\bm\bmE_{j}^\bmX)'+\bmL_{j}^\bmX R_{j}(\bmL_{j}^\bmX)'\right]\\
    \ge&\prod_{j=1}^{m}(1-\lambda_{j})\bmX
    \end{align*}
    where $\eta_{0,m}^2=\prod_{j=1}^{m}(1-\lambda_{j}),\bm\bmE_0=\bmI_n,R_0=0$ and $\bm\bmE_{j}^\bmX\mathcal{\bmT}_{j-1}(\bm\bmE_{j}^\bmX)'+\bmL_{j}^\bmX R_{j}(\bmL_{j}^\bmX)'\ge \bm0,$ $\forall j=1,2,\ldots,m.$
\item The first inequality follows straightforwardly from Fact \ref{fact 6}) above and linearity of expectation, that is,
    \begin{align*}
    \mathbb{E}\left[\mathcal{\bmM}_{m}(\bmX)\right]\ge\prod_{j=1}^{m}(1-\lambda_{j})\mathbb{E}\left[\bmX\right].
    \end{align*}
    The second inequality is due to Fact \ref{fact 5}) above which implies the concavity of the function $\mathcal{\bmM}_{m}(\bmX),$
    and therefore in the light of Jensen's inequality, it readily gives that
    \begin{align*}
    \mathcal{\bmM}_{m}\left(\mathbb{E}[\bmX]\right)\ge\mathbb{E}\left[\mathcal{\bmM}_{m}(\bmX)\right].
    \end{align*}

\end{enumerate}
\end{IEEEproof}

To take $\bm\bmP_{k+1|k}=h(\bm\bmP_{k|k})$ into consideration, the auxiliary function $\bm{\phi}_m$ can be given in the following way:
\begin{align}%
&\bm{\phi}_{-1}=h(\bmX),\bm{\phi}_{0}=h(\bmX),\nonumber\\
&\bm{\phi}_s=\sum_{j=0}^{s}\eta_{j,s}^{2}
\left(\bm\bmE_{j}\bm{\phi}_{j-1}\bm\bmE_{j}'\!+\!\bmL_{j}R_{j}\bmL_{j}'\right),s=1,2,\ldots,m\!-\!1,\nonumber\\
&\bm{\phi}_m=\sum_{j=0}^{m}\eta_{j,m}^{2}
\left(\bm\bmE_{j}\bm{\phi}_{j-1}\bm\bmE_{j}'+\bmL_{j}R_{j}\bmL_{j}'\right)\label{phi function}.
\end{align}

\begin{lemma}%
\label{varphi lemma}
Consider the function $\bm{\phi}_m(\bmL_{1},\bmL_{2},\ldots,\bmL_{m},X)$ as stated by (\ref{phi function}) with $\bmE_{j}=\bmI_n+\bmL_{j}\bmC_{j}.$  Assume $\bmX,\bmY,\bmZ\in \mathbb{S}_{+}^{n}.$ Then, the following facts hold:
\begin{enumerate}
\item With given $\bmL_{j}^{\bmX}=-\bm{\phi}_{j-1}^{\bmX}\bmC_{j}'
    \left(\bmC_{j}\bm{\phi}_{j-1}^{\bmX}\bmC_{j}'+R_{j}\right)^{-1},
    $ $j=1,2,\ldots,m,
    \bm{\varphi}(X)=\bm{\phi}_m\left(\bmL_{1}^{\bmX},
    \bmL_{2}^{\bmX},
    \ldots,\bmL_{m}^{\bmX},\bmX\right),$
    where $\bm{\phi}_{j-1}^{\bmX}=\bm{\phi}_{j-1}(\bmL_1^\bmX,\bmL_2^\bmX,\ldots,\bmL_{j-1}^\bmX,\bmX)$\label{fact31}
\item $\bm{\varphi}(X)=\min_{\bmL_{1},\bmL_{2},\ldots,\bmL_{m}}\bm{\phi}_m
    \left(\bmL_{1},\bmL_{2},\ldots,\bmL_{m},\bmX\right)
    \leq\bm{\phi}_m\left(\bmL_{1},\bmL_{2},\ldots,\bmL_{m},\bmX\right),\forall \bmL_{1},\bmL_{2},\ldots,\bmL_{m}\in\mathbb{R}^{n\times 1}$ \label{fact32}
\item If $\bmX\leq \bmY,$ then $\bm{\varphi}(\bmX)\leq \bm{\varphi}\left(\bmY\right)$ \label{fact33}
\item If $\tau\in[0,1],$ then $\bm{\varphi}\left(\tau \bmX+(1-\tau)\bmY\right)\geq \tau \bm{\varphi}(\bmX)+(1-\tau)\bm{\varphi}(\bmY)$\label{fact34}
\item $\bm{\varphi}(\bmX)\ge \prod_{j=1}^{m}(1-\lambda_{j})\bmA\bmX\bmA'+\bmQ$\label{fact35}
\item If $\widebar{\bmX}\ge \bm{\varphi}(\widebar{\bmX}),$ then $\widebar{\bmX}>{\bm0}$ \label{fact36}
\item For a random variable $\bmX,$ $\prod_{j=1}^{m}(1-\lambda_{j})\bmA{\mathbb{E}}[\bmX]\bmA'+\bmQ
    \leq {\mathbb{E}}\left[\bm{\varphi}(\bmX)\right]\leq \bm{\varphi}\left({\mathbb{E}}\left[\bmX\right]\right).$\label{fact37}
\end{enumerate}
\end{lemma}
\begin{IEEEproof}
We only prove Fact \ref{fact36}) because the others can be derived directly from Lemma \ref{m lemma}.{\color{black}{

\begin{enumerate}
\item[6)] According to Fact \ref{fact37}) above, it gives that $\widebar{\bmX}\ge\bm{\varphi}\big(\widebar{\bmX}\big)\ge
    \prod_{j=1}^{m}(1-\lambda_{j})\bmA\widebar{\bmX}\bmA'+\bmQ.$ Since $\left(\bmA,\bmQ^{\frac{1}{2}}\right)$ is controllable, then there must exist an $\hat{\bmX}>{\bm0}$ subject to the Lyapunov equation $\hat{\bmX}=\prod_{j=1}^{m}(1-\lambda_{j})\bmA\hat{\bmX}\bmA'+\bmQ$ if $\sqrt{\prod_{j=1}^{m}(1-\lambda_{j})}\bmA$ is asymptotically stable. Accordingly, it follows that
    \begin{align*}
    \widebar{\bmX}-\hat{\bmX}>\prod_{j=1}^{m}(1-\lambda_{j})\bmA(\widebar{\bmX}-\hat{\bmX})\bmA'
    \end{align*}
    implying there exists a $\hat{\bmQ}>{\bm0}$ such that
    \begin{align*}
    \widebar{\bmX}-\hat{\bmX}=\prod_{j=1}^{m}(1-\lambda_{j})\bmA(\widebar{\bmX}-\hat{\bmX})\bmA'+\hat{\bmQ}.
    \end{align*}
    Therefore, $\widebar{\bmX}-\hat{\bmX}>{\bm0},$ or $\widebar{\bmX}>\hat{\bmX}>{\bm0}.$ This completes the proof.
\end{enumerate}
}}
\end{IEEEproof}

\begin{remark}%
\label{p bound}
Observe that if we substitute $\bmX={\color{black}{\bmP_{k|k}}}$ into Fact \ref{fact37}) in Lemma \ref{varphi lemma}, it follows that $\prod_{j=1}^{m}(1-\lambda_{j})\bmA{\mathbb{E}}\left[\bmP_{k}\right]\bmA'+\bmQ\leq{\mathbb{E}}\left[\bm{\varphi}(\bmP_{k})\right]
\leq\bm{\varphi}\left({\mathbb{E}}\left[\bmP_{k}\right]\right).$ Since ${\mathbb{E}}\left[\bmP_{k+1}|\bmP_{k}\right]=\bm{\varphi}(\bmP_{k})$ and ${\mathbb{E}}\left[\bmP_{k+1}\right]={\mathbb{E}}\left[\bm{\varphi}(\bmP_{k})\right],$ then $\prod_{j=1}^{m}(1-\lambda_{j})A{\mathbb{E}}\left[\bmP_{k}\right]A'+Q\leq{\mathbb{E}}\left[\bmP_{k+1}\right]\leq\bm{\varphi}\left({\mathbb{E}}\left[\bmP_{k}\right]\right).$ That is, the expected value of $\bmP_{k+1|k}$ can be lower-bounded and upper-bounded by $\prod_{j=1}^{m}(1-\lambda_{j})\bmA{\mathbb{E}}\left[\bmP_{k}\right]\bmA'+\bmQ$ and $\bm{\varphi}\left({\mathbb{E}}\left[\bmP_{k}\right]\right)$ both as functions of ${\mathbb{E}}\left[\bmP_{k}\right],$ respectively.
\end{remark}

To facilitate the convergence analysis, let us define the linear part of function $\bm{\phi}_m(\bmL_{1},\bmL_{2},\ldots,\bmL_{m},\bmX)$ in terms of variable $\bmX$ as another auxiliary function, namely
\begin{align}
\mathcal{L}_m(\bmY)=\sum_{j=0}^{m}\eta_{j,m}^{2}
\left(\bmE_{j}\bm{\phi}_{j-1}\bmE_{j}'\right)\label{l function}
\end{align}
where $\bm{\phi}_{j-1},j=0,1,\ldots,m$ are defined in (\ref{phi function}). Then, the following lemma can be readily presented.
\begin{lemma}%
\label{l lemma}
Consider the function $\mathcal{\bmL}_m(\bmY)$ as stated in (\ref{l function}). If there exists a positive definite matrix $\widebar{\bmY}>{\bm0}$ such that $\widebar{\bmY}>\mathcal{\bmL}_m(\widebar{\bmY}),$ then
\begin{enumerate}
\item $\forall \bmW\ge {\bm0},$ $\lim_{k\to\infty}\mathcal{\bmL}_m^{k}(\bmW)={\bm0}$
\item Given $\bmU>{\bm0},$ let the following sequence
\begin{align*}%
\bmY_{k+1}=\mathcal{\bmL}_m(\bmY_{k})+\bmU
\end{align*}
initialized at $\bmY_{0}\ge \bm0.$ Then, the sequence $\bmY_{k}$ is bounded.
\end{enumerate}
\end{lemma}
\begin{IEEEproof}%
\begin{enumerate}
\item Note that $\mathcal{\bmL}_m(\bmY)$ is affine in $\bmY$ and $\mathcal{\bmL}_m(\bmY)\ge {\bm0}, \forall \bmY\ge {\bm0},$ and $\mathcal{\bmL}_m(\bmY)\geq \mathcal{\bmL}_m(\bmZ), $ for $ \bmY\ge\bm \bmZ.$ There exist constants $0\leq r<1$ and $t\ge 0$ such that $\mathcal{\bmL}_m(\widebar{\bmY})\le r\widebar{\bmY}<\widebar{\bmY}$ and $\bmW\leq t\widebar{\bmY},$ respectively. Then
\begin{align}%
0\leq \mathcal{\bmL}_m^{k}(\bmW)\leq t\mathcal{\bmL}_m^{k}(\widebar{\bmY})\le tr^{k}\widebar{\bmY}.\label{ineq}
\end{align}
Therefore, it can be readily obtained that ${\bm0}\leq\lim_{k\to\infty}\mathcal{L}_m^{k}(\bmW)\le\lim_{k\to\infty}tr^{k}\widebar{\bmY}\to {\bm0}$ given that $0\le r<1.$
\item {\color{black}{Based on (\ref{ineq}) above, for any initialization $\bmY_0\ge {\bm0}$ and any $\bmU>{\bm0},$  there always exist two constants $t_{\bmY_{0}}\ge 0$ and $t_{\bmU}\ge 0$ such that $\bmY_0\le t_{\bmY_0} \widebar{\bmY}$ and $\bmU\le t_\bmU\widebar{\bmY},$ which are independent of $k.$ }}Therefore, similar arguments in (\ref{ineq}) lead to
\begin{align*}%
\bmY_{k}=&\mathcal{\bmL}_m^{k}(\bmY_{0})+\sum_{s=0}^{k-1}\mathcal{\bmL}_m^{s}(\bmU)\\
\leq&{\color{black}{t_{\bmY_{0}}}}r^{k}\widebar{\bmY}+\sum_{s=0}^{k-1}t_{\bmU}r^{s}\widebar{\bmY}\\
=&\left({\color{black}{t_{\bmY_{0}}}}r^{k}+t_{\bmU}\frac{1-r^{k}}{1-r}\right)\widebar{\bmY}.
\end{align*}
Obviously, the result on the boundedness of the sequence $Y_{k}$ holds true.
\end{enumerate}
\end{IEEEproof}

\begin{lemma}%
\label{phi convergence}
Consider the function $\bm{\phi}_m(\bmL_{1},\bmL_{2},\ldots,\bmL_{m},\bmX)$ defined in (\ref{phi function}). Assume there exist $m$ gain matrices $\widebar{\bmL}_{1},\widebar{\bmL}_{2},\ldots,\widebar{\bmL}_{m}$ and
a positive definite matrix $\widebar{\bmP}$ such that
\begin{align*}%
\widebar{\bmP}>{\bm0}~\text{and}~\widebar{\bmP}>\bm{\phi}_m
\left(\widebar{\bmL}_{1},\widebar{\bmL}_{2},\ldots,\widebar{\bmL}_{m},\widebar{\bmP}\right).
\end{align*}
Then, the sequence $\bmP_{k}=\bm{\varphi}^{k}(\bmP_{0})$ is bounded for any given $\bmP_{0}.$ That is, there exists a positive definite matrix $\bmM_{\bmP_{0}}>{\bm0}$ depending on $\bmP_{0}$ such that
$$\bmP_{k}\leq \bmM_{\bmP_{0}},\forall k\ge 0.$$
\end{lemma}
\begin{IEEEproof}%
Observe that $\bm{\phi}_m(\bmL_{1},\bmL_{2},\ldots,\bmL_{m},\bmY)=\mathcal{\bmL}_m(\bmY)+\bmQ+\mathcal{\bmN}_m,$ where $\mathcal{\bmN}_m:=\sum_{j=0}^{m}\eta_{j,m}^{2}\left(\bmE_j\mathcal{\bmN}_{j-1}\bmE_j'+\bmL_jR_j\bmL_j'\right)\ge{\bm0}$ with $\mathcal{\bmN}_{0}={\bm0},$ $\bmQ\ge {\bm0},$ and $R_{j}\ge 0,j=0,1,\ldots,m.$ Therefore,
\begin{align*}%
\widebar{\bmP}&>\bm{\phi}_m(\widebar{\bmL}_{1},\widebar{\bmL}_{2},\ldots,\widebar{\bmL}_{m},\widebar{\bmP})\\
&=\mathcal{\bmL}_m(\widebar{\bmP})+\bmQ+\mathcal{\bmN}_m\\
&\ge \mathcal{\bmL}_m(\widebar{\bmP}).
\end{align*}
That is, $\widebar{\bmP}> \mathcal{\bmL}_m(\widebar{\bmP}),$ hence, the function $\mathcal{\bmL}_m(\bmY)$ satisfies the condition of Lemma \ref{l lemma}. Considering the definition of $\bm{\varphi}(\bmP_{k}),$ it yields that
\begin{align*}%
\bmP_{k+1}=\bm{\varphi}(\bmP_{k})&\leq \bm{\phi}_m(\bmL_{1},\bmL_{2},\ldots,\bmL_{m},\bmP_{k})\\
&=\mathcal{\bmL}_m(\bmP_{k})+\bmQ+\mathcal{\bmN}_m\\
&=\mathcal{\bmL}_m(\bmP_{k})+\bmU
\end{align*}
where $\bmU:=\bmQ+\mathcal{\bmN}_m\ge {\bm0}.$ Then based on fact 2) in Lemma \ref{l lemma}, it can be concluded that the sequence $\bmP_{k}$ is bounded for any $k\ge 0$.
\end{IEEEproof}
\begin{lemma}%
\label{lemma hh}
Let $\bmY_{s+1}=f(\bmY_{s})$ and $\bmZ_{s+1}=f(\bmZ_{s}).$ Suppose that the function $f(\bmY)$ is monotonically increase in $\bmY.$ Then:
\begin{align*}
\bmY_{1}&\ge \bmY_{0}\Longrightarrow \bmY_{s+1}\ge \bmY_{s},\forall s\ge 0\\
\bmY_{1}&\le \bmY_{0}\Longrightarrow \bmY_{s+1}\le \bmY_{s},\forall s\ge 0\\
\bmY_{0}&\le \bmZ_{0}\Longrightarrow \bmY_{k}\le \bmZ_{s},\forall s\ge 0.
\end{align*}
\end{lemma}
\begin{IEEEproof}%
The three statements can be similarly proved by mathematical induction. Thus, due to page limitation, we here only prove the first one. Since $\bmY_{1}\ge \bmY_{0},$ then the first statement is true for $k=0.$ Then assume that $\bmY_{t+1}\ge \bmY_{t}$ holds, so $\bmY_{t+2}=f(\bmY_{t+1})\ge f(\bmY_{t})=\bmY_{t+1}$ holds owing to the monotonicity of function $f(\bmY).$
\end{IEEEproof}

{\color{black}{
After building these lemmas above, we are now in a position to establish the sufficient condition for mean square stability of the averaged estimation error covariance matrix.

\begin{theorem}[Sufficient condition]
\label{sufficient condition}
Consider the function $\bm{\phi}_m=\sum_{j=0}^{m}\eta_{j,m}^{2}
\big(\bmE_{j}\bm{\phi}_{j-1}\bmE_{j}'+\bmL_{j}R_{j}\bmL_{j}'\big)$ defined in (\ref{phi function}).
If there exist $m$ matrices $\tilde{\bmL}_{j},j=1,2,\ldots,m$ and a positive definite matrix $\tilde{\bmP}$ such that
\begin{align}%
\tilde{\bmP}>{\bm0}~\text{and}~\tilde{\bmP}>\bm{\phi}_m(\tilde{\bmL}_{1},
\tilde{\bmL}_{2},\ldots,\tilde{\bmL}_{m},\tilde{\bmP}).\label{cond}
\end{align}
Then, the following facts are true:
\begin{enumerate}%
\item The MARE converges for any initial condition $\bmP_{0}\ge {\bm0}$ and the limit
$$\lim_{t\to\infty}\bmP_{k}=\lim_{k\to\infty}\bm{\phi}_m^{k}(\bmP_{0})=\widebar{\bmP}$$
is independent of the initial condition $\bmP_{0}.$
\item $\widebar{\bmP}$ is the unique positive definite fixed point of the MARE.
\end{enumerate}
\end{theorem}
}}

\begin{IEEEproof}%
1) To begin with, we verify the convergence of the MARE sequence initialized at $\bmQ_0={\bm0}$ and therefore $\bmQ_k=\bm{\varphi}^k(\bm0).$  Then it directly follows that ${\bm0}=\bmQ_0\le\bm{\varphi}(\bm0)=\bmQ_1,$ and in the light of Fact \ref{fact33} in Lemma \ref{varphi lemma}, it gives that
\begin{align*}
\bmQ_1=\bm{\varphi}(\bmQ_0)\leq \bm{\varphi}(\bmQ_1)=\bmQ_2.
\end{align*}
From Lemma \ref{lemma hh} and according to Lemma \ref{phi convergence}, a monotonically nondecreasing sequence of matrices follow straightforwardly from a simple inductive argument and the sequence is also upper-bounded, that is,
\begin{align*}
{\bm0}=\bmQ_0\le \bmQ_1\le \bmQ_2\le\ldots\le \bmM_{\bmQ_0}.
\end{align*}
Here, one can easily verify that the monotonically nondecreasing and upper-bounded sequence converges from the Bolzano-Weierstrass theorem, that is,
\begin{align*}
\lim_{k\to\infty}\bmQ_k=\widebar{\bmP}
\end{align*}
where $\widebar{\bmP}\ge {\bm0}$ is a fixed point of the following modified Riccati iteration
\begin{align}
\widebar{\bmP}=\bm{\varphi}\big(\widebar{\bmP}\big).\label{barp}
\end{align}

Then, we show that the modified Riccati iteration initialized at $\bmS_0\ge \widebar{\bmP}$ also converges to the same point $\widebar{\bmP}.$ By resorting to (\ref{l function}), it gives that
\begin{align*}
\widebar{\bmP}=\bm{\varphi}\big(\widebar{\bmP}\big)&=\mathcal{\bmL}_m^{{\widebar{\bmP}}}\big(\widebar{\bmP}\big)+\bmQ+\mathcal{\bmN}_{m}^{\widebar{\bmP}}\\
&>\mathcal{\bmL}_m^{{\widebar{\bmP}}}\big(\widebar{\bmP}\big)
\end{align*}
where $\mathcal{\bmL}_m^{{\widebar{\bmP}}}(\bmY)=\sum_{j=0}^{m}\eta_{j,m}^2 \left[\bmE_{j}^{{\widebar{\bmP}}}\bm{\phi}_{j-1}\left(\bmE_{j}^{{\widebar{\bmP}}}\right)'\right].$ Therefore, the function $\mathcal{\bmL}_m^{\widebar{\bmP}}$ satisfies the condition of Lemma \ref{l lemma}. Accordingly, we realize that
\begin{align*}
\lim_{k\to\infty}\big(\mathcal{\bmL}_m^{\widebar{\bmP}}\big)^k(\bmY)={\bm0},\quad\forall \bmY\ge {\bm0}.
\end{align*}
Assume that $\bmS_0\ge \widebar{\bmP}$ and then,
\begin{align*}
\bmS_1=\bm{\varphi}\big(\bmS_0\big)\ge \bm{\varphi}\big(\widebar{\bmP}\big)=\widebar{\bmP}
\end{align*}
where is due to the monotonically increase property of the function $\bm{\varphi}(\bmX)$ and (\ref{barp}). By induction, it establishes that
\begin{align*}
\bmS_k\ge \widebar{\bmP},\quad\forall k>0.
\end{align*}
Meanwhile, we have
\begin{align*}
{\bm0}\le& \bmS_{k+1}-\widebar{\bmP}=\bm{\varphi}\big(\bmS_k\big)-\bm{\varphi}\big(\widebar{\bmP}\big)\\
=&\bm{\phi}_m\big(\bmL_1^{\bmS_{k}},\bmL_2^{\bmS_{k}},\ldots,\bmL_m^{\bmS_{k}},\bmS_{k}\big)-\bm{\phi}_m\big(\bmL_1^{\widebar{P}},\bmL_2^{\widebar{\bmP}},\ldots,\bmL_m^{\widebar{\bmP}},\widebar{\bmP}\big)\\
\le&\bm{\phi}_m\big(\bmL_1^{\widebar{\bmP}},\bmL_2^{\widebar{\bmP}},\ldots,\bmL_m^{\widebar{\bmP}},\bmS_{k}\big)-\bm{\phi}_m\big(\bmL_1^{\widebar{\bmP}},\bmL_2^{\widebar{\bmP}},
\ldots,\bmL_m^{\widebar{\bmP}},\widebar{\bmP}\big)\\
=&\sum_{j=0}^{m}\eta_{j,m}^2\Big[\bmE_{j}^{\widebar{\bmP}}\left(\bm{\phi}_{j}^{\bmS_k}-\bm{\phi}_{j}^{\widebar{\bmP}}\right)\big(\bmE_{j}^{\widebar{\bmP}}\big)'\Big]\\
=&\mathcal{\bmL}_m^{\widebar{\bmP}}\big(\bmS_k-\widebar{\bmP}\big).
\end{align*}
Then, since $\lim_{k\to\infty}\mathcal{L}_m^{\widebar{\bmP}}\big(\bmS_k-\widebar{\bmP}\big)= {\bm0},$ it directly follows that $\lim_{k\to\infty}(\bmS_{k+1}-\widebar{\bmP})={\bm0}.$
That is, we have shown ${\color{black}{\bmS_k}}\to\widebar{\bmP}$ as $k\to\infty$ when $\bmS_0\ge\widebar{\bmP}.$

In the following, we are ready to justify that the modified Riccati iteration $\bmP_k=\bm{\varphi}^k(\bmP_0)$ converges to $\widebar{\bmP}$ for all initial conditions $\bmP_0\ge {\bm0}.$ Let $\bmQ_0={\bm0}$ and $\bmS_0=\widebar{\bmP}+\bmP_0.$ Then consider the three Riccati iterations initialized at $\bmQ_0,\bmP_0$ and $\bmS_0,$ respectively. Clearly, $\bm\bmQ_0\le \bm\bmP_0\le \bm\bmS_0,$ and in the light of Lemma \ref{lemma hh}, it gives that ${\bm0}\le \bmQ_k\le \bmP_k\le \bmS_k,~ \forall k\ge 0.$
Given that both the sequence $\bmQ_k$ and the sequence $\bm\bmS_k$ converge to $\widebar{\bmP},$ consequently, we have $\lim_{k\to\infty}\bm\bmP_k=\widebar{\bmP}.$

2) Let us further postulate there exists another positive semi-definite matrix $\hat{\bmP}\ge {\bm0}$ such that $\hat{P}=\bm{\varphi}\big(\hat{\bmP}\big).$ Let us consider the Riccati iteration initialized at $\hat{\bmP},$ and therefore, we can derive the following sequence
\begin{align*}
\hat{\bmP},\hat{\bmP},\hat{\bmP},\ldots.
\end{align*}
From analysis above, it has been shown that every Riccati iteration converges to the same limit $\widebar{\bmP}.$ Therefore, we have $\hat{\bmP}=\widebar{\bmP}.$
\end{IEEEproof}

{\color{black}In the sequel, we will provide an example of a scalar-state vector-observation system to justify the existence of sufficient condition in Theorem \ref{sufficient condition}.

{\emph{Example:}} We consider the following system 
\begin{align*}
x_{k+1}=&ax_k+\omega_k\\
\bmy_k=&\bmC x_k+\bmupsilon_k,
\end{align*}
where $a=1.2,$ $\bmC'=[c_1,c_2]=[1,1],$ noise covariances are $q=1$ and $\bmR={\rm{diag}}\{r_1,r_2\}={\rm{diag}}\{0.1,1\}.$ For simplicity, consider $\lambda_1=\lambda_2=0.6,$ and let $l_1,l_2$ be, for instance, such that $l_1=-1,-2.8276< l_2<0.8276$ or $l_2=-1,-2.8276<l_1<0.8276.$ Then one can always find $p>0$ such that $l_1,l_2,p$ satisfy condition (\ref{cond}) in Theorem \ref{sufficient condition}. That is, the expected estimation error covariance matrix will converge.}

{\color{black}{
In the ensuing part, we will present one necessary condition for ensuring mean square stability of expected estimation error covariance matrix which extends the result in \cite{tsp2012you} to general linear systems with data packet drops.

\begin{theorem}[Necessary condition]%
\label{necessary condition}
Consider system (\ref{system}) and Algorithm \ref{ps}. Assume that $\bmA$ is unstable, that $\left(\bmA,\bmQ^{1/2}\right)$ is controllable and that $(\bmC,\bmA)$ is observable. If ${\rm{E}}[\bm\bmP_{k}]\leq \bmM_{\bm\bmP_{0}}, \forall k\ge 0$ holds for any initial condition $\bm\bmP_{0}\ge {\bm0},$ then $\lambda_{1},\lambda_{2},\ldots,\lambda_{m}$ defined in (\ref{lambda}) should satisfy the following condition
\begin{align}%
\label{my condition}
\prod_{i=1}^{m}(1-\lambda_{i})\leq \frac{1}{\Big(\max_{i} \left|\sigma_{i}(\bmA)\right|\Big)^{2}}
\end{align}
where {\color{black}{$\sigma_i(\bmA), i=1,2,\ldots,n,$ are all eigenvalues of square matrix $\bmA,$}} and $\bmM_{\bm\bmP_{0}}>{\bm0}$ depends on the initial condition $\bm\bmP_{0}\ge{\bm0}.$
\end{theorem}
}}

\begin{IEEEproof}
The proof follows straightforwardly from Fact \ref{fact37}) in Lemma \ref{varphi lemma}.
\end{IEEEproof}

\section{Concluding Remarks}
\label{section 4}
In this paper we devised a measurement innovation componentwise based power scheduler for wireless sensors in terms of optimally deciding whether to use a high or low transmission power to communicate each component of a measurement to the remote estimator side. The high transmission power is used to transmit the well-defined ``important'' measurements and low transmission power to transmit the less ``important'' measurements. Meanwhile, the high power transmission power is assumed to lead to reliable data flow while the low transmission power leads to unreliable data flow, that is, data packet drops. Under this new framework, the MMSE estimator was derived. Then convergence analysis of the averaged estimation error covariance was provided and moreover, both the sufficient condition and necessary condition guaranteeing its convergence were established for general linear stochastic systems. {\color{black}{Since the assumption of modeling the arrival of measurements as independent Bernoulli i.i.d. processes can be clearly improved upon, and therefore, future work will concentrate on accounting for communication channel modeling in this filtering framework \cite{auto2009dey, tsp2006luo}.}}


\section*{Acknowledgment}
The authors would like to express thanks to the anonymous reviewers for their insightful and constructive comments
that helped improving the quality of this paper.

%


\end{document}